\documentclass{aa}  

\usepackage{xcolor}
\usepackage{graphicx}
\usepackage{txfonts}
\usepackage{siunitx}
\usepackage{natbib}
\usepackage{amsmath}
\usepackage{subfigure}
\usepackage{hyperref}
\hypersetup{
    colorlinks=true,
    citecolor=blue,
    linkcolor=blue,
    urlcolor=blue,
}

\usepackage{orcidlink}
\newcommand{\orcid}[1]{\href{https://orcid.org/#1}
{{\includegraphics[height=8pt]{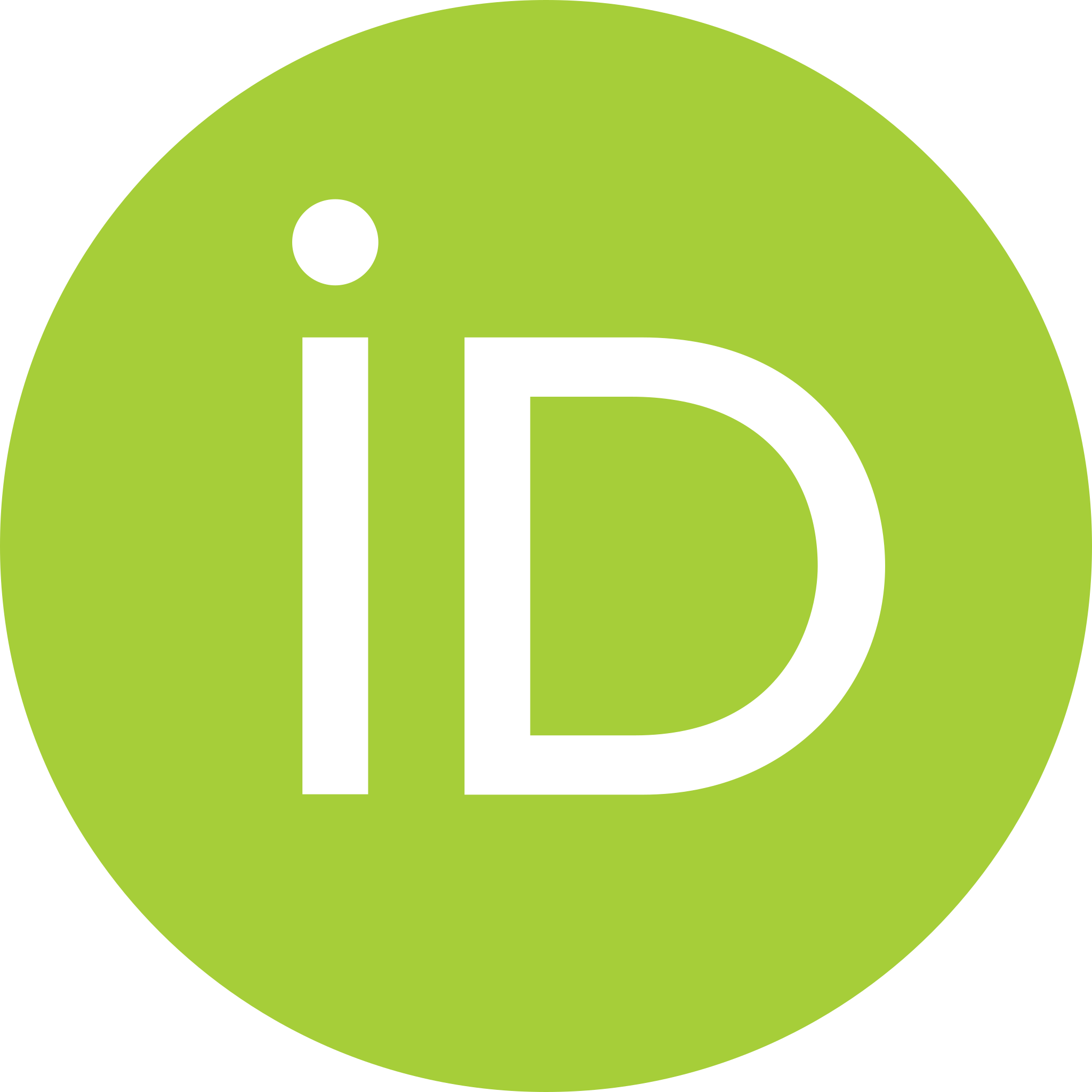}}}}

\def\bhm{M$_{\rm BH}$}
\def\lbhm{log(M$_{\rm BH})$}
\def\msolar{M$_{\odot}$}
\def\lum{L$_{\rm bol}$}
\def\llum{log(L$_{\rm bol}$)}
\def\ledd{$\lambda_{Edd}$}
\def\lledd{log($\lambda_{Edd})$}
\def\lambdarest{$\lambda_{\rm rest}$}
\def\psdamp{PSD$_{\rm amp}$}

\def\psdslope{PSD$_{\rm slope}$}

\usepackage{ulem}   
\definecolor{mpc}{rgb}{0.2, 0.2, 0.8}
\definecolor{vpc}{rgb}{0.2, 0.8, 0.2}

\begin{document}

   \title{Ensemble power spectral density of SDSS quasars in UV/optical bands}

    \author{V. Petrecca\thanks{vincenzo.petrecca@unina.it}\inst{1,2}\orcid{0000-0002-3078-856X}
          \and
          I. E. Papadakis\inst{3,4}\orcid{0000-0001-6264-140X}
          \and
          M. Paolillo\inst{1,2,5}\orcid{0000-0003-4210-7693}
          \and 
          D. De Cicco\inst{1,2,7}\orcid{0000-0001-7208-5101}
          \and
          F. E. Bauer\inst{6,7,8}\orcid{0000-0002-8686-8737}
          }

   \institute{Dipartimento di Fisica    "Ettore Pancini", Università di Napoli Federico II, via Cinthia 9, 80126 Napoli, Italy
        \and 
        INAF - Osservatorio Astronomico di Capodimonte, Salita Moiariello 16, 80131 Napoli, Italy
        \and
        Department of Physics and Institute of Theoretical and Computational Physics, University of Crete, 71003 Heraklion, Greece
        \and
        Institute of Astrophysics, FORTH, GR-71110 Heraklion, Greece
        \and
        INFN — Sezione di Napoli, via Cinthia 9, 80126, Napoli, Italy
        \and
        Instituto de Astrof{\'{\i}}sica and Centro de Astroingenier{\'{\i}}a, Facultad de F{\'{i}}sica, Pontificia Universidad Cat{\'{o}}lica de Chile, Campus San Joaquín, Av. Vicuña Mackenna 4860, Macul Santiago, Chile, 7820436
        \and
        Millennium Institute of Astrophysics, Nuncio Monse{\~{n}}or S{\'{o}}tero Sanz 100, Of 104, Providencia, Santiago, Chile
        \and
        Space Science Institute, 4750 Walnut Street, Suite 205, Boulder, Colorado 80301
        }

   \date{Received 4 January 2024 / Accepted 8 April 2024}

 
  \abstract
   {Quasar variability has proven to be a powerful tool to constrain the properties of their inner engine and the accretion process onto supermassive black holes. Correlations between UV/optical variability and physical properties have been long studied with a plethora of different approaches and time-domain surveys, although the detailed picture is not yet clear.}
   {We analysed archival data from the SDSS Stripe-82 region to study how the quasar power spectral density (PSD) depends on the black hole mass, bolometric luminosity, accretion rate, redshift, and rest-frame wavelength. We developed a model-independent analysis framework that could be easily applied to upcoming large surveys such as the Legacy Survey of Space and Time (LSST).}
   {We used light curves of 8042 spectroscopically confirmed quasars, observed in at least six yearly seasons in five filters $ugriz$. We split the sample into bins of similar physical properties containing at least 50 sources, and we measured the ensemble PSD in each of them.}
   {We find that a simple power law is a good fit to the power spectra in the frequency range explored. Variability does not depend on the redshift at a fixed wavelength. Instead, both PSD amplitude and slope depend on the black hole mass, accretion rate, and rest-frame wavelength. We provide scaling relations to model the observed variability as a function of the physical properties, and discuss the possibility of a universal PSD shape for all quasars, where frequencies scale with the black hole mass, while normalization and slope(s) are fixed (at any given wavelength and accretion rate).}
   {}


   \maketitle
%
\section{Introduction}
\label{sec:intro}

Variability is one of the most striking features of quasars and has been known since they were discovered (\citealt{1963ApJ...138...30M}). The last 60 years of observations have shown that it is observed at all wavelengths, both in the continuum and the emission lines, with timescales ranging from minutes to years and a completely stochastic and aperiodic behaviour (e.g. \citealt{Ulrich1997}; \citealt{Giveon1999}; \citealt{Hawkins2002}; \citealt{VandenBerk+2004}; \citealp{Wilhite2005,Wilhite2008}). In effect, X-ray and UV/optical power spectral densities (PSDs, or power spectra) extracted from light curves of quasars show no periodic features and a typical red-noise trend, usually modelled with one or more power laws (e.g. \citealt{Collier-Peterson2001}; \citealt{Uttley2002}; \citealt{McHardy04}). 

Quasars are the most luminous among non-obscured active galactic nuclei (AGN). It is generally accepted that most of their UV/optical light comes from thermal emission attributed to an accretion disc, while X-rays are produced by inverse-Compton scattered photons from a hot electron corona surrounding the central supermassive black hole (SMBH; see \citealt{Netzer2015} and references therein). However, the exact interplay between the two phenomena, as well as the origin of the variability itself, is still debated. Some authors support the idea of a thermal origin with propagation of instabilities through the accretion disc and/or changes in the accretion rate (e.g. \citealt{Neustadt2022} and references therein). Other evidences favour the idea of optical variability due to X-ray reprocessing (e.g. \citealt{Kammoun21}; \citealt{Panagiotou22}), but a combination of multiple effects is also not excluded. 

The last two decades have seen an increase in both the quality and quantity of multi-epoch UV/optical data from large ground-based surveys, such as the Optical Gravitational Lensing Experiment (OGLE; \citealt{Kelly+09}; \citealt{Kozlowski2010}; \citealt{Zu+13}), the Sloan Digital Sky Survey (SDSS) Stripe-82 region (\citealt{Sesar2006}; \citealp{MacLeod10,MacLeod2012}; \citealt{Guo+17}; \citealt{Yu2022}), the Palomar-QUEST Survey (\citealt{Bauer2009}), the Panoramic Survey Telescope And Rapid Response System survey (PanSTARRS-1; \citealt{Morganson2014}; \citealt{Simm2016}), LaSilla-Quest (\citealt{Cartier2015}; \citealt{SanchezSaez2018}), the VST SUDARE-VOICE (\citealt{Falocco2015}; \citealp{DeCicco2015,DeCicco2019,DeCicco2022}; \citealt{Poulain2020}), the Palomar Transient Factory (PTF/iPTF surveys; \citealt{Caplar2017}), the Hyper Suprime-Cam Subaru Strategic Program (\citealt{Kimura2020}), the Catalina Real-time Transient Survey (\citealt{Rakshit2017}; \citealt{Graham2020}; \citealt{Tachibana2020}; \citealt{Laurenti2020}), the NASA Asteroid Terrestrial-impact Last Alert System (ATLAS; \citealt{Tang2023}), and the Zwicky Transient Facility (ZTF; \citealt{LopezNavas2023}; \citealt{Arevalo2023_1}). Space missions such as Kepler and the Transiting Exoplanet Survey Satellite (TESS) also provided data with a high cadence (e.g. \citealt{Mushotzky+11}; \citealt{Edelson2014}; \citealt{Kasliwal+15}; \citealt{Treiber2023}). 

Many efforts in combining data from different surveys enabled the variability on long timescales to be constrained. \cite{Rumbaugh2018} joined SDSS and Dark Energy Survey (DES) light curves, \cite{Suberlak+21} used SDSS Stripe-82 and PanSTARRS-1 data, while \cite{Stone2022} studied a sample of quasars observed with SDSS, PanSTARRS-1, DES, and a dedicated monitoring with the DECam on the CTIO-4m Blanco telescope, over 20 years. The near future will be even more promising. The Vera C. Rubin Observatory's Legacy Survey of Space and Time (LSST; \citealt{LSSTbook}; \citealt{Ivezic2019}), expected to start in 2025, will observe the entire southern sky every 3--4 nights for 10 years, across the $ugrizy$ filters. This new facility will detect tens of millions of AGN (also using variability as a selection tool, see \citealt{Savic2023}), allowing a thorough analysis of light curves, including fast variability (e.g. \citealt{Raiteri2022}) and precise time-delay measurements (\citealt{Kovacevic2022}; \citealt{Czerny2023}). Tests to define the final cadence requirements are still ongoing at the time of writing (e.g. \citealt{Brandt2018}; \citealt{Kovacevic2021}; \citealt{Sheng2022}).

All surveys mentioned above allow us to study the statistical properties of optical variability and its correlations with the physical parameters of quasars. Ensemble analyses have shown that rest-frame variability amplitude increases with time (i.e. the red-noise trend of power spectra). Variability is also observed to be anti-correlated with luminosity (and/or the accretion rate) at all timescales, and many works also find an anti-correlation with rest-frame wavelength. Dependencies on the black hole mass, \bhm, are less certain, as well as the overall shape of the PSD (i.e. spectral indices and characteristic timescales). A partial list comparing the observed correlations between variability and quasar properties is shown in Table 4 of \cite{Suberlak+21}, along with the different methods adopted to measure variability. 

Optical light curves are often modelled using a damped random walk (DRW; \citealt{Kelly+09}), a first-order continuous-time autoregressive moving-average (CARMA) process. DRW is a statistical model which describes AGN variability via an exponential covariance matrix with two parameters: a variability amplitude, $\sigma$, and a characteristic damping timescale, $\tau$. Such a model predicts a power-law PSD with a spectral index of --2, flattening to 0 (i.e. white noise) for times $>> \tau$. However, DRW processes are sometimes studied in time-domain via the structure function (SF), defined as the root mean square (rms) magnitude difference as a function of the time difference $\Delta t$ between observation pairs (\citealt{Kozlowski2016} and references therein). DRW (or associated SF parameters) can be then studied as a function of physical properties of quasars (e.g. \citealt{MacLeod10}; \citealt{Zu+13}; \citealt{Kasliwal+15}; \citealt{Suberlak+21}). Although there are known biases in the recovery of reliable damping timescales from observed light curves, as well as uncertainties on the SF itself due to sampling effects (e.g. \citealt{Emmanoulopoulos2010}; \citealt{kozlowski+17}), many attempts have been made to relate the DRW $\tau$ to any physical timescale (see e.g. \citealt{Burke+21}). Regardless of the proper estimate of $\tau$, variability amplitude predicted by the DRW model seems to be consistent with light curves length of the order of a few years, but deviations have been observed on both longer and shorter timescales (e.g. \citealt{Mushotzky+11}; \citealt{Guo+17}). Possible steps forward may invoke the usage of higher-order CARMA processes, such as the damped harmonic oscillator (DHO; \citealt{Yu2022}), indirect PSD estimation through CARMA modelling (\citealt{Kelly2014}), or unsupervised machine learning analysis of time series (\citealt{Tachibana2020}).

In this work, we propose a model-independent variability analysis of $\sim 9000$ spectroscopically confirmed quasars from the SDSS Stripe-82 region through direct PSD measurement. We show how to account for irregular sampling and take advantage of the large dataset to execute ensemble analysis and constrain the PSD shape as a function of redshift, luminosity, black hole mass, accretion rate, and wavelength. We provide scaling relations for the PSD on timescales of $\sim$ [250,1500] days (rest-frame) and present a framework which could be easily applied to other large multi-epoch surveys. We will discuss the PSD shape on shorter and longer timescales in a future work (Petrecca et al. in prep.).

This paper is organized as follows. In Section \ref{sec:thesample} we present the data used for our work. Section \ref{sec:method} describes the variability measurement method through PSD, while Section \ref{sec:results} is dedicated to the analysis of the correlations between variability and physical properties. Finally, we summarize our results and discuss possible interpretation and future perspectives in Section \ref{sec:summary}.

\section{The sample}
\label{sec:thesample}
Our analysis requires a large sample of AGN with very well-known properties and well-sampled light curves. Among all the surveys listed in the Introduction, the Stripe-82 region of the SDSS \citep{2000AJ....120.1579Y} suits our purposes perfectly. This area consists of $290 \ deg^2$ with right ascension from 20:00h to 4:00h and declination from $-1.26^\circ$ to $+1.26^\circ$, imaged multiple times by the SDSS from 2000 to 2008. There are $\sim 60$ observations in 5 filters \textit{u,g,r,i,z} (\citealt{1996AJ....111.1748F}; \citealt{1998AJ....116.3040G}; \citealt{2002AJ....123.2121S}) grouped in yearly seasons about 2–3 months long, albeit most of the data are in the last 6 seasons. Images in the different filters are nearly simultaneous, separated by $71.72$ seconds of drift scan time.

\begin{figure}
    \includegraphics[width=1.0\columnwidth]{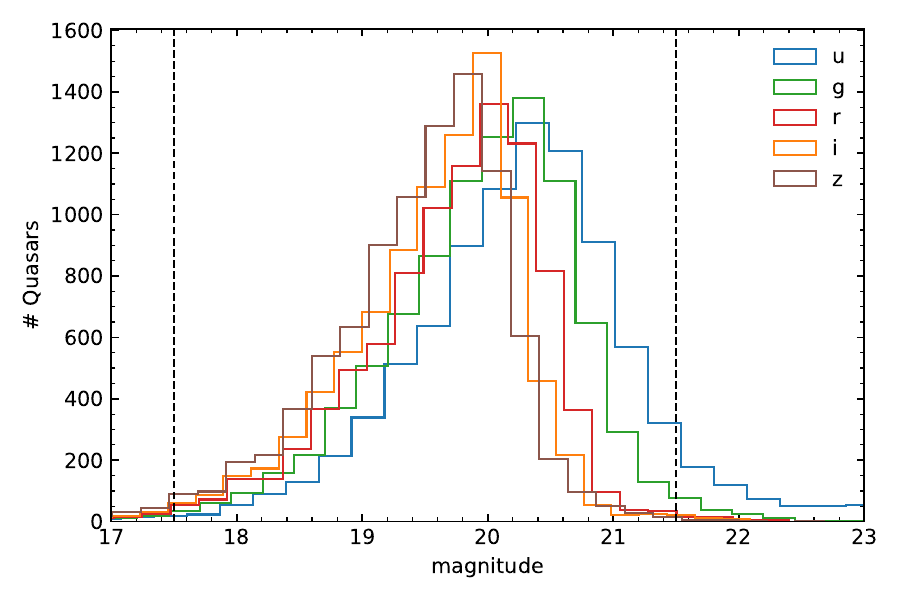}
    \caption{Distribution of observed quasar magnitudes in the five SDSS filters. Dashed lines enclose the magnitude range of sources used in this analysis.}
    \label{fig:magsdistr}
\end{figure}

Our sample of AGN consists of $9275$ well-characterized, spectroscopically confirmed quasars selected by \citet{MacLeod10,MacLeod2012}\textbf, who provide tables with light curves.\footnote{\hyperlink{https://faculty.washington.edu/ivezic/macleod/qso_dr7/index.html}{https://faculty.washington.edu/ivezic/macleod/qso\_dr7/index.html}} We retrieved virial black hole masses and bolometric luminosities for 9148 quasars from the SDSS DR7 quasar catalogue by \cite{2011ApJS..194...45S}. Masses are estimated from emission line widths ($H_\beta$ for $z < 0.7$, Mg {\small{II}} for $0.7 \le z< 1.9$, and C {\small{IV}} for $z \ge 1.9$). The nominal mean uncertainty on fiducial mass estimates is estimated to be $\sim 0.2$ dex, although systematics from reverberation mapping could be as high as $\sim 0.5$ (\citealt{Krolik2001}). Despite such uncertainties, the wide range of available masses still allows us to study whether variability depends on mass, as explained in Sect. \ref{sec:results}. Bolometric luminosities are derived from the continuum using bolometric corrections derived by \cite{2006ApJS..166..470R} from the composite spectral energy distribution ($BC_{5100}=9.26$ for $z < 0.7$, $BC_{3000}=5.15$ for $0.7 \le z< 1.9$, and $BC_{1350}=3.81$ for $z \ge 1.9$). Average uncertainties on bolometric luminosities are of the order of 0.1 dex. We discard from the sample 125 sources with no valid estimate of black hole mass. Figure \ref{fig:magsdistr} shows the distribution of the quasar magnitudes in the \textit{ugriz} bands. We restricted the sample to magnitudes $17.5 < mag_{\rm AB} < 21.5$ in all bands, as there are few quasars with $mag_{\rm AB} < 17.5$, while at $mag_{\rm AB} > 21.5$ there are no stars available for the necessary computation of the Poisson noise (see Sect. \ref{sec:method}).

\begin{figure*}
    \includegraphics[width=0.33\linewidth]{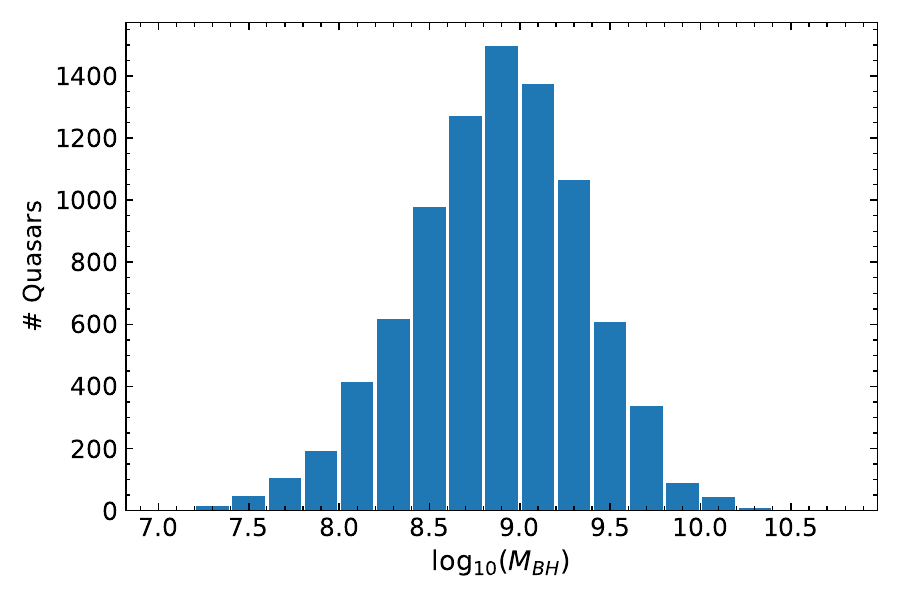}
    \includegraphics[width=0.33\linewidth]{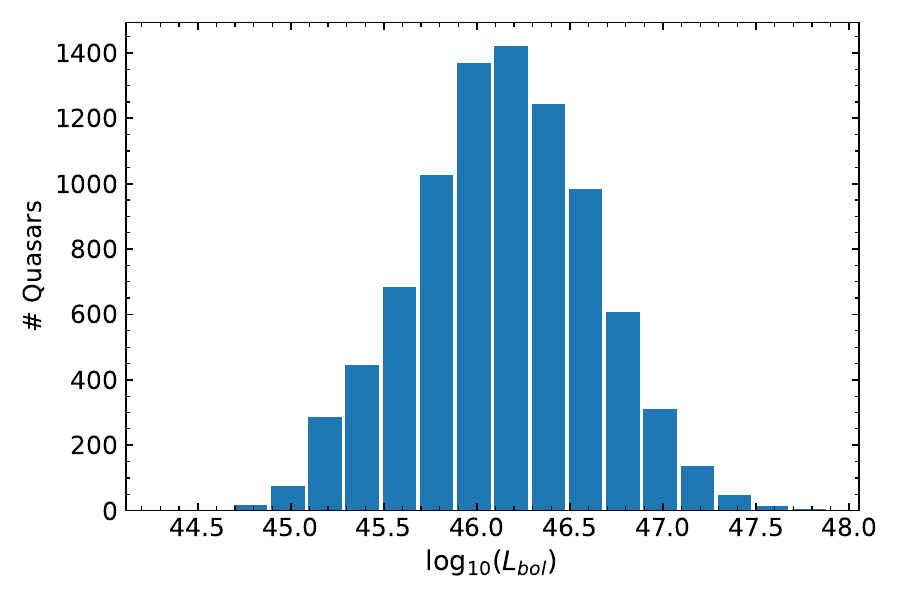}
    \includegraphics[width=0.33\linewidth]{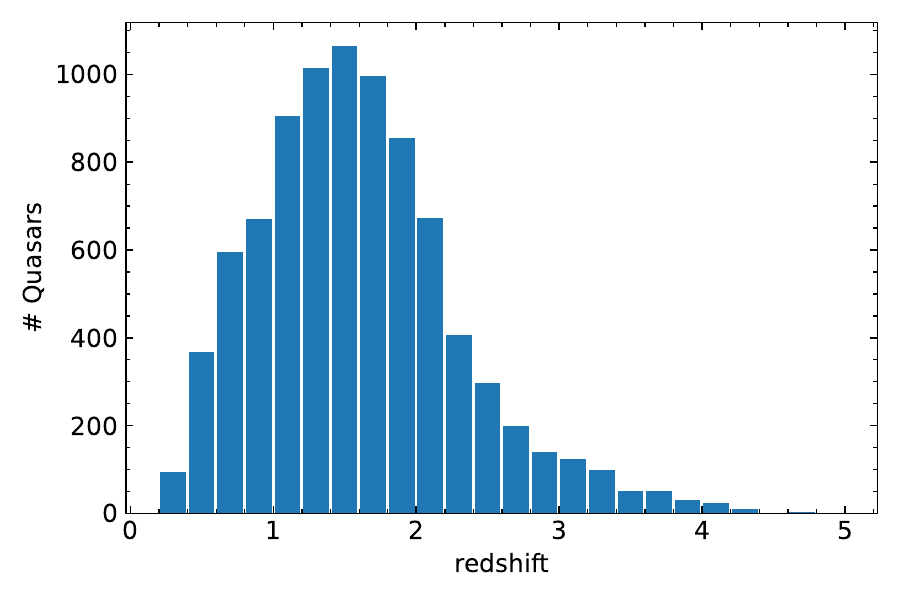}
    \caption{Distribution of black hole mass, bolometric luminosity, and redshift for the quasars in our sample.}
    \label{fig:quasarpropdistr}
    \end{figure*}

The sample, after the magnitude cuts, consists of 8119 quasars with known black hole mass, \bhm, bolometric luminosity, \lum, and redshift, $z$. Their distribution is shown in Fig. \ref{fig:quasarpropdistr} (left, middle and right panels, respectively). We cover a broad range of masses (from $\sim 10^{7.5}$ up to $\sim 10^{10}$ \msolar), and luminosities (10$^{45}$ -- 10$^{47}$ ergs s$^{-1}$), which are typical of the quasars. Most of the sources have redshifts $0.5 \leq z \leq 2.5$, although a tail at larger redshifts is also apparent in the figure. The large number of quasars over a broad range of \bhm\ and \lum\ values will give us the opportunity to study if and how the variability properties of quasars (quantified by the PSD analysis) depend on each of the two parameters. The distribution of a large range of redshift will also allow us to investigate if the quasar optical variability properties evolve with time. 

\begin{figure}
    \includegraphics[width=1.0\columnwidth]{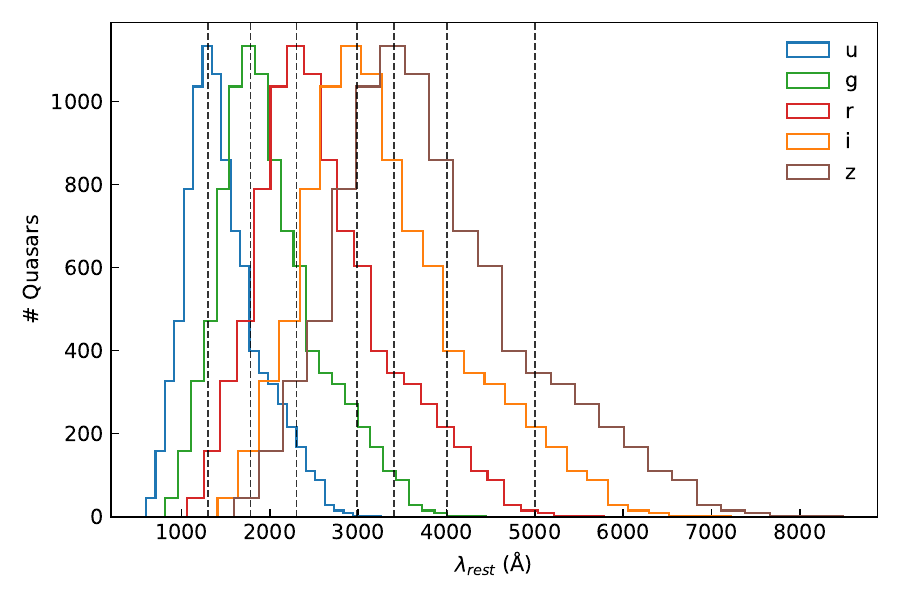}
    \caption{Distribution of rest-frame wavelengths (\lambdarest) of all light curves, for all quasars in the sample. The vertical dashed lines indicate the centres of the \lambdarest\ bins at which we computed the PSD of quasars.}
    \label{fig:restlambda}
\end{figure}

In addition, the availability of light curves across five bands will enable us to compare variability properties over different rest-frame continuum wavelengths. Figure \ref{fig:restlambda} shows the distribution of the rest-frame wavelength of each light curve in all filters, of all quasars, computed as $\lambda_{rest}=\lambda_{obs}/(1+z)$, where $\lambda_{obs,u}=3520\AA$, $\lambda_{obs,g}=4800\AA$, $\lambda_{obs,r}=6250\AA$, $\lambda_{obs,i}=7690\AA$, and $\lambda_{obs,z}=9110\AA$ are the SDSS photometric system central wavelengths. The five left-most vertical dashed lines in the figure indicate the wavelength where the \textit{u,g,r,i} and \textit{z} rest-frame wavelength distribution peak is located (\lambdarest$_{,i}$ = 1300, 1800, 2300, 3000, and 3400\AA, where $i=1,2,3,4$ and $5$). The other two vertical lines indicate two additional rest-frame wavelengths, namely \lambdarest$_{,6}$ = 4000 and \lambdarest$_{,7}$ = 5000\AA. 
When we compute the power spectrum of quasars in various rest-frame bands (see Sect. \ref{sec:results}), we consider all light curves, irrespective of the filter, whose rest-frame wavelength is within $|\Delta\lambda|/$\lambdarest = 0.2, around the \lambdarest\ wavelengths indicated by the vertical lines in Fig. \ref{fig:restlambda}. As for each quasar we have five distinct light curves in \textit{ugriz}, we only use the one whose \lambdarest\ is the closest to the centre of the selected wavelength bin. The number of quasars within the \lambdarest$_{,6}$, and in particular within the \lambdarest$_{,7}$ band, is smaller than the number of quasars in the other rest-frame wavelength bins. We considered those longer wavelengths, so that we can study the energy dependence of the power spectrum over of range of \lambdarest$_{,7}$/\lambdarest$_{,1}\sim 4$ in energy separation.

As this work makes use of broad-band photometry data, we expect some contamination from the broad line region (BLR) to the accretion disc continuum (mostly broad lines, blended iron multiplets, and the Balmer pseudo-continuum). Although spectroscopic studies of AGN from the SDSS show contamination levels up to $\sim 20\%$ in the most severe cases, the mean contribution over a broad wavelength region is typically less than $10\%$ (\citealt{Vandenberk2001}; \citealt{Calderone2017}). Besides, the light-crossing time for the BLR is one order of magnitude greater than for the UV/optical disc, on average (\citealt{Cackett2021}). For this reason, we do not expect a major contribution from line variability to our analysis. Host-galaxy contamination is also negligible, as it starts to be significant for low-luminosity quasars with $z \lesssim 0.5$ and at \lambdarest\ above $5000 \AA$ (\citealt{2011ApJS..194...45S}), while we mostly sample higher-luminosity and higher-redshift quasars at bluer wavelengths.

\section{The variability analysis method}
\label{sec:method}
\subsection{The SDSS light curves}

\begin{figure*}
    \includegraphics[width=0.5\linewidth]{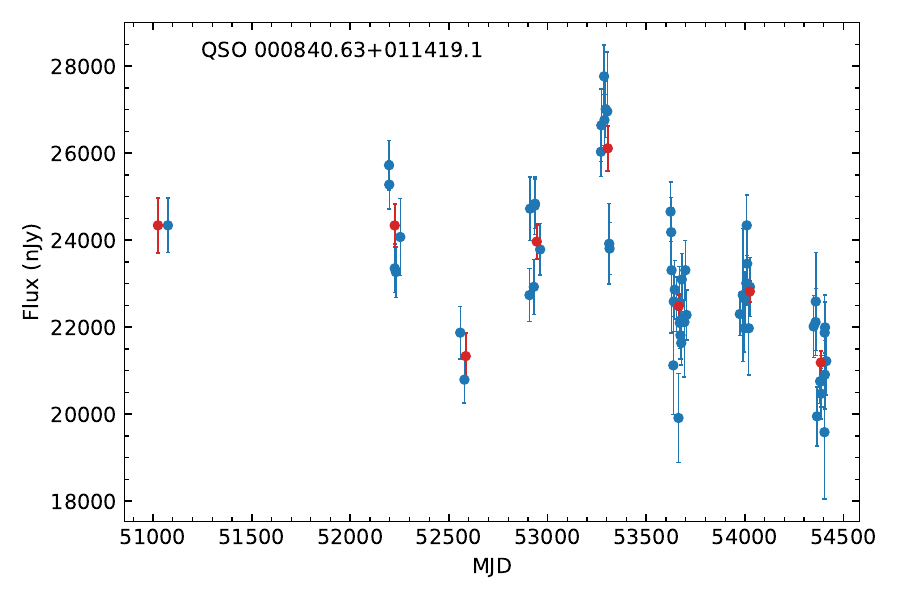}
    \includegraphics[width=0.5\linewidth]{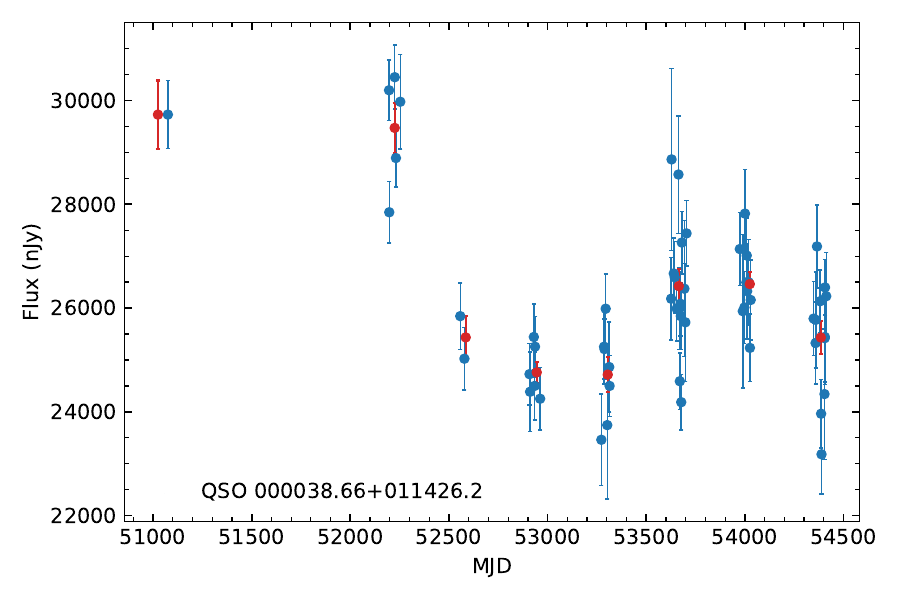}
    \caption{Light curves of two quasars in g-band with a similar BH mass (8.8 < log(\bhm) < 9.0), luminosity (46.1 < \llum \ < 46.3), and redshift ($1.1<z<1.3$). Red points indicate the binned light curve.}
    \label{fig:lc}
\end{figure*}

Stripe-82 was observed for almost 10 years, with a denser sampling cadence as the survey progressed, yielding very similar light curves in terms of observed sampling for all quasars in our sample. As an example, Fig. ~\ref{fig:lc} shows the g-band light curves of two quasars, which have similar \bhm, \lum\ and redshift. In the later observing seasons, sources were observed approximately 10 times in $\sim 3$ months. SDSS magnitudes have been corrected for galactic extinction (coefficients from \citealt{schlegel98}) and converted into fluxes in units of nJy, assuming an AB photometric system with a zero point of $3631 Jy$ and following a standard procedure for error propagation.

 We computed the (straight) mean flux per observing season (red dots in Fig. \ref{fig:lc}). The observing seasons are equally spaced by 1 year in the observer frame. We verified that the deviations w.r.t. the mean epoch of all observations are of the order of just a few sigma. Errors on the binned fluxes are computed using the standard error on the mean $\sigma/\sqrt{n}$, where $\sigma$ is the standard deviation and $n$ is the number of points in each season. In this way we obtain evenly sampled light curves, with at least seven points. This is the case for almost all quasars in our sample, given the similarity of Stripe-82 observations in terms of cadence for all sources.

The second season data point at around $\sim 52200$ MJD is missing in Fig. \ref{fig:lc} as well as in many quasar light curves. In addition, there are a few more missing points among the last six observations for 77 sources. For that reason, we rejected them and kept $N_Q=8042$ sources in the sample, for which we can compute binned light curves, with no missing points in the last six observations, like the light curves shown in Fig.~\ref{fig:lc}.

\subsection{Power spectrum derivation}
\label{sec:psdestimation}
We used the last six seasons of the binned light curves, where points are evenly sampled, and we computed the periodogram of each one, that is,
\begin{equation}
I_{N}(f_j)  =  \frac{2\Delta t}{N\Bar{x}^2} \ \Bigg[\sum^{N}_{i=1} [x(t_i)-\Bar{x}]e^{-i2\pi f_jt_i}\Bigg]^2,
\label{periodogram}
\end{equation}
\noindent which we accept as an estimate of the intrinsic power spectrum. In the equation above, $N=6$ is the number of points in the binned light curves used for the analysis, $x(t_i)$ are the binned flux values, and $\Bar{x}$ is the mean flux. We computed the rest-frame power spectrum of each source, that is we assume $\Delta t=365/(1+z)$ is the size of the time window, and all times $t_i$ are divided by $(1+z)$. Frequencies are defined as $f_j=j/(N\Delta t)$, with $j=1,2$ and 3 (by using six points in the binned light curves, we end up with three frequencies). 

The periodogram is the most common estimator of the PSD function of a variable process. The PSD is the Fourier transform of the autocovariance function of the process, (which holds all information about the important variability properties of the process, including the presence of intrinsic timescales) and it is frequently used in many research fields, including astronomy (e.g. \citealt{Press1978}; \citealt{Priestley1981}; \citealt{vanderklis1988}; and references therein). Knowledge of the intrinsic PSD of a variable source can be very important in constraining various physical (or statistical) models that have been proposed to interpret the observed variations of an AGN in the UV/optical bands. Although power spectral analysis is a well known and used tool for time series interpretation, it may suffer from biases depending on the sampling pattern and duration; thus we further validated the robustness of our method via simulations, as shown in Appendix \ref{app_B}.

Observed PSDs should also be affected by the experimental Poisson noise. In principle, we could predict the resulting Poisson noise level knowing the photometric accuracy of the data. This is not true in our case due to various reasons. For example, as we mentioned above, we need to retrieve the original uncertainties transforming back the magnitudes to fluxes, but this process is usually based on a first-order approximation.  In addition, there could be residual uncertainties (e.g. calibration) that are not properly taken into account when estimating the error on magnitudes. We therefore followed a different approach, and we computed the Poisson noise level using the light curves of non-variable stars that have the same magnitudes as the quasars in our sample, as suggested by, e.g. \cite{kozlowski+16}, and as we explain in detail in Appendix \ref{app_A}.

Figure \ref{fig:psdexample_single} show the rest-frame g-band PSDs, using the binned light curves plotted in Fig.\,~\ref{fig:lc}. Since we consider only the last six points in the binned light curves, we compute the power spectrum in 3 frequencies, over half a decade in frequency space. 
Solid lines indicate the best-fit straight line to the PSDs (in the log-log space). The line appears to be a reasonable fit to the data, but we cannot compute errors on the best-fit parameters and we cannot judge the goodness of fit, since we do not know the error of each PSD measurement, as it depends on the (unknown) intrinsic PSD (e.g. \citealt{1993MNRAS.261..612P}). 
However, we can do much better than just estimating the PSD of each source in the sample, over three frequencies, by taking advantage of the large number of quasars in the sample. For example, we can group the quasars into narrow black hole (BH) mass, luminosity and redshift bins, compute the (noise-corrected) periodogram of each one of them, and then average all the periodograms in order to compute the ensemble power spectrum of all quasars in each [\bhm, \lum, $z$] bin. 

\begin{figure}
    \begin{center}
    \includegraphics[width=1.0\columnwidth]{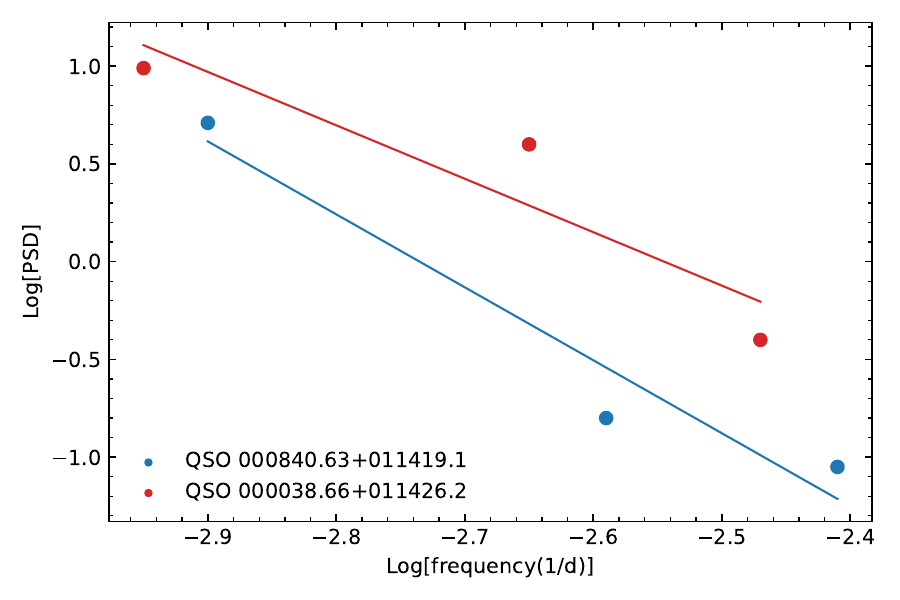}
    \end{center}
    \caption{Noise-corrected PSD of the two light curves shown in Fig. \ref{fig:lc}. The solid lines indicate the best-fit line to the data.}
    \label{fig:psdexample_single}
\end{figure}

\begin{figure}
    \begin{center}
    \includegraphics[width=1.0\columnwidth]{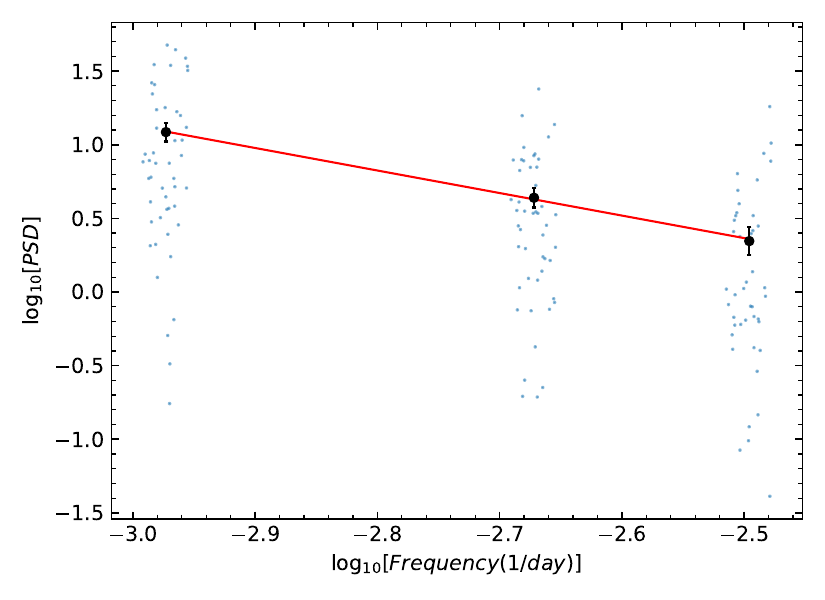}
    \end{center}
    \caption{Logarithm of the noise-corrected periodogram of all quasars with [8.8 < \lbhm \ < 9.0], [46.1 < \llum \  < 46.3], and [1.2 < $z$ < 1.4]. Small blue dots indicate single PSD estimates, while black points show the logarithm of the mean PSD (with its error), and the red line is the best-fit linear model to the ensemble PSD.}
    \label{fig:psdexample_ensemble}
\end{figure}

We show the logarithm of the (noise-corrected) periodogram of all quasars which belong to the BH mass, luminosity and redshift bin with: [8.8 < \lbhm \ < 9.0], [46.1 < \llum \ < 46.3], and [1.2 < $z$ < 1.4] in Fig.\,~\ref{fig:psdexample_ensemble}. The periodograms are calculated in the rest-frame of each source and, since we consider a redshift bin with a small width of 0.2, the difference in rest-frame frequencies is small. We therefore compute the mean of all the periodograms in each frequency bin, and the standard error on the mean, and we accept this as representative of the average power spectrum of quasars with the chosen BH mass and luminosity. As long as there are at least 50 objects in each bin (and hence 50 periodogram estimates at each frequency), the mean periodogram should approximately follow a Gaussian distribution (\citealt{1993MNRAS.261..612P}). In practice, we compute the logarithm of the mean periodogram, and its error (following the standard error propagation formula), as it is more convenient to fit straight lines to sampled power-spectra, when the intrinsic power spectrum follows a power-law form (as is usually the case with the quasar power spectra).  

We fitted the mean PSD  in Fig. \ref{fig:psdexample_ensemble} with a straight line, in log--log space, i.e.
\begin{equation}
\log_{10}[{\rm PSD}(\nu)]=\log_{10}[{\rm PSD_{amp}]+PSD_{slope}}[\log(\nu)+2.6],
\label{eq:psdmod}
\end{equation}

\noindent where \psdamp\, is the power spectrum amplitude at $\nu=10^{-2.6}$ day$^{-1}$, i.e. $\sim 400$ days. We chose to normalize the PSD at this frequency, as it is always within the range of sampled frequencies for our sample. In this way the error of the best-fit parameters is minimized. The solid red line in Fig. \ref{fig:psdexample_ensemble} shows the best-fit line to the mean PSD ($\chi^2=0.05$ for 1 degree of freedom/dof). 

\section{Power spectral analysis results}
\label{sec:results}

As we already mentioned, our main objective is to study the optical power spectrum of quasars, using the light curves of 8042 sources in Stripe-82 in five bands \textit{ugriz}. The main physical parameters of AGN are the BH mass, \bhm, and the accretion rate (derivable from the bolometric luminosity through the Eddington ratio, \ledd),\footnote{The ratio of the bolometric Luminosity, \lum, and the Eddington Luminosity.} which determine the emitted energy spectrum. Thus, irrespective of the mechanism that drives the optical variability, it is plausible that the variability properties may also depend on these physical parameters or, perhaps, remain constant in all quasars. In addition, given the fact that quasars are powered by accretion of gas onto the BH, we expect the energy released through the accretion process, as well as the disc temperature, to increase inwards. This implies that the outer parts of the disc will be able to emit only longer wavelength photons. As the variability mechanism may depend on the local disc properties, we may expect the variability properties to also depend on radius, and hence, on the energy of the emitted photons (or their wavelength, i.e. \lambdarest).

Based on the discussion above, we grouped the quasars into bins of BH mass and luminosity-accretion rate, we computed their average PSD (as explained in Sect. \ref{sec:psdestimation}) in various rest-frame wavelengths, we fitted it with the model defined by Eq.\, \ref{eq:psdmod}, we determined \psdamp, and \psdslope, and we investigated whether \psdamp\, and \psdslope\ depend on \bhm, \ledd, and \lambdarest, or if they are simply the same in all quasars, irrespective of energy and the quasar physical properties. Other relevant properties of AGN variability might be the line-of-sight orientation, the radio-loudness, the presence of a jet, and the BH spin, but these are much harder to measure. However, our sample consists only of quasars, i.e. with likely similar orientation, while significantly different variability amplitudes are only expected for extreme radio-loud sources ($< 1\%$ of the sample, see Table 2 in \citealt{MacLeod10}). 

We define BH mass bins, of size 0.2 dex, in the range 8.2 < \lbhm \ < 9.6, and \lum\ bins of the same size, in the range 45.3 < \llum \ < 47.0. The lower/upper limits as well as the width of the bins are determined by the fact that, on the one hand, we need large bins with as many objects as possible but, on the other hand, objects in each bin should also have (almost) the same \bhm\ and \lum. The first constraint requires large bin sizes, while the second  implies the choice of small bin sizes. Given the  uncertainty on the \bhm\ measurements (from the width of emission lines measured in optical spectra and systematics from reverberation mapping studies) and on estimating \lum (from bolometric correction factors), a choice of a bin size of $0.2$ may include some objects with similar BH mass and luminosity (and hence \ledd) in adjacent bins. Nonetheless, the wide range of masses and luminosities allows us to define many different bins and study any PSD trend with physical properties of quasars. We also define rest-frame wavelength (i.e. rest-frame energy) bins. The centres of those bins are indicated by the vertical lines plotted in Fig. \ref{fig:restlambda}. As we explained in Sect. \ref{sec:thesample}, the width of each wavelength bin is such that, $|\Delta$\lambdarest$|/$\lambdarest = 0.2. 

We compute the PSD of each single quasar, and then the mean PSD of all quasars in each [\bhm, \ledd(or \lum), \lambdarest] bin (as we explained in Sect. \ref{sec:psdestimation},) and we fitted it in the log--log space using the straight line model defined by Eq.~\ref{eq:psdmod}. The number of quasars is larger than 50 in all bins (to ensure proper $\chi^2$ statistics when assessing the goodness of fit to the PSDs). We found that a simple line can fit the mean quasar PSDs, in all bins. We discuss below how the best-fit PSD amplitude and slope depend on the various parameters. But before doing this, we first investigate the possibility that the variability properties may depend on redshift, which would imply that the variability mechanism may evolve with time.

\subsection{Dependence of PSD on redshift}
\label{sec:redshift}

\begin{figure}
    \begin{center}
    \includegraphics[width=1.0\columnwidth]{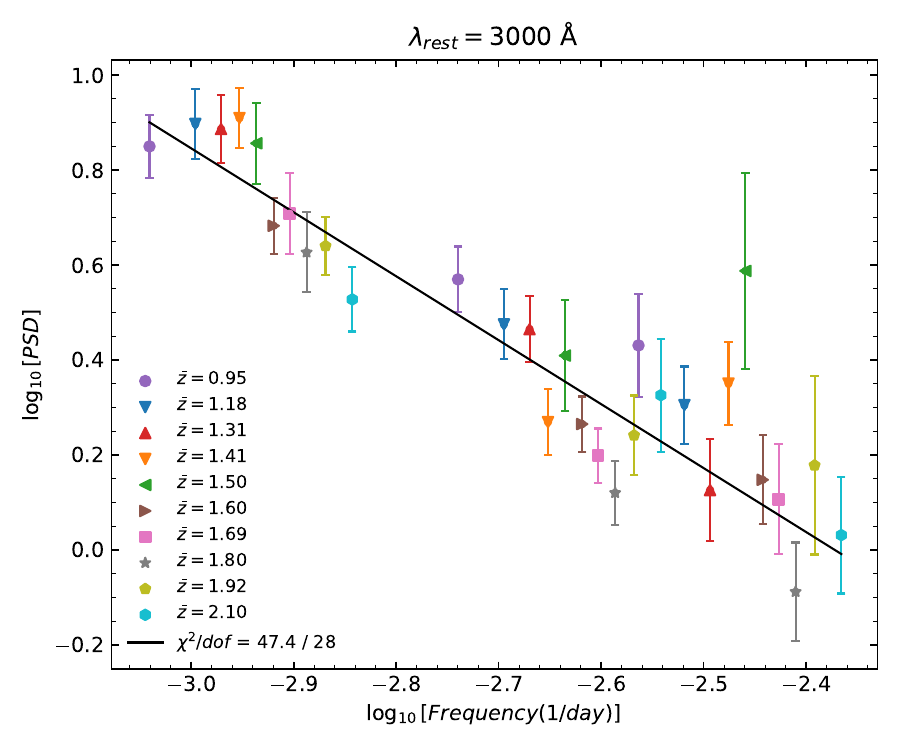}
    \end{center}
    \caption{Ensemble PSD of quasars with 8.8 < \lbhm \ < 9.1 and 46.0 < \llum \ < 46.3 at \lambdarest = 3000 \AA \ in ten redshift bins (different symbols and colours). The average redshift for each bin is reported in the legend.}
    \label{fig:zcor}
\end{figure}

In order to investigate whether the PSD varies with redshift, we considered quasars with [8.8 < \lbhm \ < 9.0] and [46.1 < \llum \ < 46.3]. This BH mass and luminosity bin is the one with the largest number of quasars (see the peak of the BH mass and luminosity distribution in Fig.\, \ref{fig:magsdistr}) and has a slightly larger bin size of 0.3 to further maximize the number of sources at different redshift. 
For each \lambdarest, we divided the quasars in this [\bhm,\lum] bin into various redshift bins, and we computed the mean PSD in each one of them (as described in Sect. \ref{sec:psdestimation}). The results in the case when we use light curves with \lambdarest=3000\AA\ are shown in Fig. \ref{fig:zcor}. The redshift bins were chosen such that each bin included at least 50 objects (to ensure Gaussian statistics, as we explain in Sect. \ref{sec:psdestimation}). The average redshift in each bin is listed in the legend in Fig.\, \ref{fig:zcor}, and their widths range from $\Delta z=0.05$ to $\sim 0.2-0.3$, in a few cases.  

We fitted all the PSDs plotted in Fig. \ref{fig:zcor} at all redshifts, from $z\sim 0.9-2$, with the model defined by Eq. \ref{eq:psdmod}. The solid line shows the resulting best-fit with a straight line.
The best-fit $\chi^2$ value is $\chi^2=47.4$ for 28 dof, and the resulting $p_{null}$\footnote{The $p_{null}$ is the probability to obtain a value equal or greater than our test statistic from random data extracted from the best-fit model.} is 0.012. Formally speaking, this means that we cannot reject (at the 99\% confidence level) the null hypothesis that all points are drawn from the same PSD, and hence we can assume that the PSD is independent of redshift. In any case, even if there were some residual dependence on redshift, it would not be large enough to explain other strong trends which we report below. 

Although we just report the results for \lambdarest = 3000\AA \ as an example, we computed power spectra using light curves at all the rest frame wavelengths defined in Sect. \ref{sec:thesample} for the same \bhm\ and \lum\ bin defined above, obtaining the same results. Furthermore, we repeated the same exercise considering quasars in smaller/larger BH mass/luminosity bins, so that we could sample smaller and larger redshift ranges, from $z\sim 0.5$ to 2.5. In all cases, we found that the power spectra of quasars at different redshift bins are consistent with the same power-law model. This implies that we do not find significant evidence for the dependence of the quasar optical variability power spectrum on redshift. For the rest of the analysis, we consider all quasars with redshift from 0.5 to 2.5, when we compute their mean PSD in various mass and luminosity bins.

\subsection{Dependence of PSD on \lum} 

\begin{figure}
    \includegraphics[width=1.0\columnwidth]
    {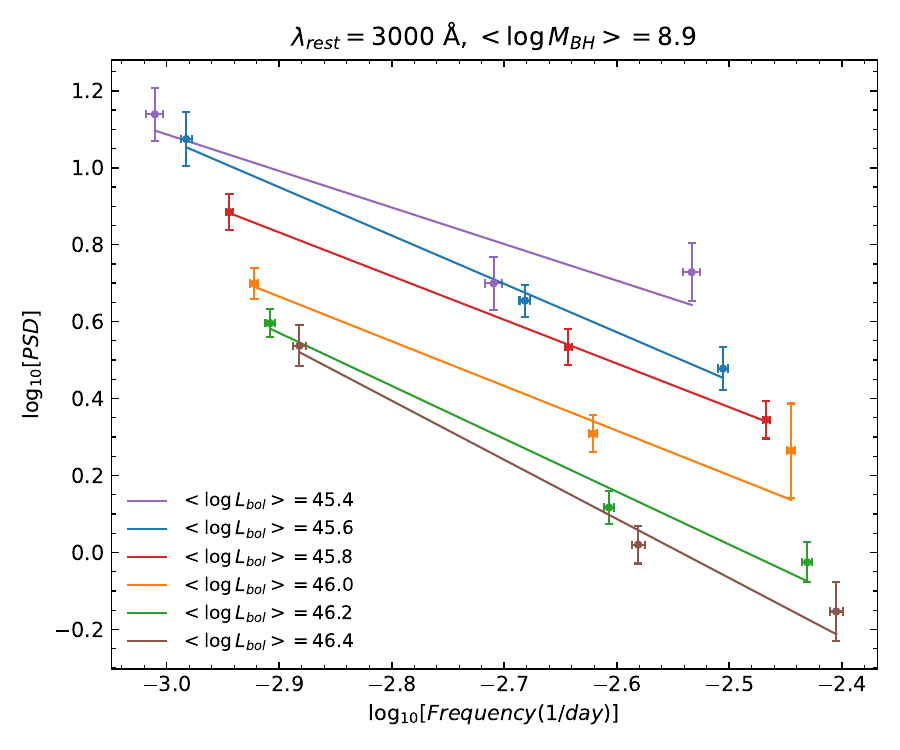}
    \caption{Ensemble PSDs of quasars with 8.8 < \lbhm \ < 9.0 and all redshifts between 0.5 and 2.5 in different \llum \ bins. The legend reports the average luminosity in each bin of width 0.2.}
    \label{fig:psd_lum}
\end{figure}

 We proceed to investigate the dependence of the power spectrum on the bolometric luminosity, \lum. Figure \ref{fig:psd_lum} shows the mean PSD of all quasars with [8.8 < \lbhm \ < 9.0] (i.e. the BH mass bin with the largest number of quasars, see left panel in Fig. \ref{fig:quasarpropdistr}), in six luminosity bins from \llum = 45.3 up to 46.5. The luminosity bin size is $\Delta$\llum = 0.2, and we have used the \lambdarest = 3000\AA\, light curves to compute the PSDs.  The solid lines show the best-fit model lines (as defined by Eq.\,~\ref{eq:psdmod}). The model fits the data rather well. Since there are just three PSD points in each luminosity bin, there are some issues with the resulting error on the best-fit model parameter values.   
 In some cases, the three PSD points just happen to be located exactly on the best-fit line (i.e. the red points in specific case of Fig.~\ref{fig:psd_lum}, where \llum = 45.8), while in other cases the points are spread around the best fit (e.g. the purple points in the same panel, with \llum = 45.4). Consequently, the error on the best-fit parameters varies considerably. Given the small number of data points in each PSD, the error from the best model-fits to the individual PSDs may not be representative of the true error of the best-fit parameters. For that reason, in all cases below, we consider the mean of all the best-fit parameter errors as a more representative error of the respective best-fit parameter.

 Figure \ref{fig:psd_lum} shows that the power spectrum depends strongly on the source luminosity. The power spectra at different luminosity bins appear to have similar slopes but different amplitudes (we discuss possible mild dependencies of the slopes later in the text). It appears that the PSD amplitude increases with decreasing luminosity. This result is in agreement with a trend that has been known for a long time in many quasars, at all wavebands: more luminous sources are less variable (e.g. \citealt{Barr1986}; \citealt{Cristiani1996}; \citealt{1993MNRAS.261..612P}).

\subsection{Dependence of PSD on \bhm}
\begin{figure}
    \includegraphics[width=1.0\columnwidth]
    {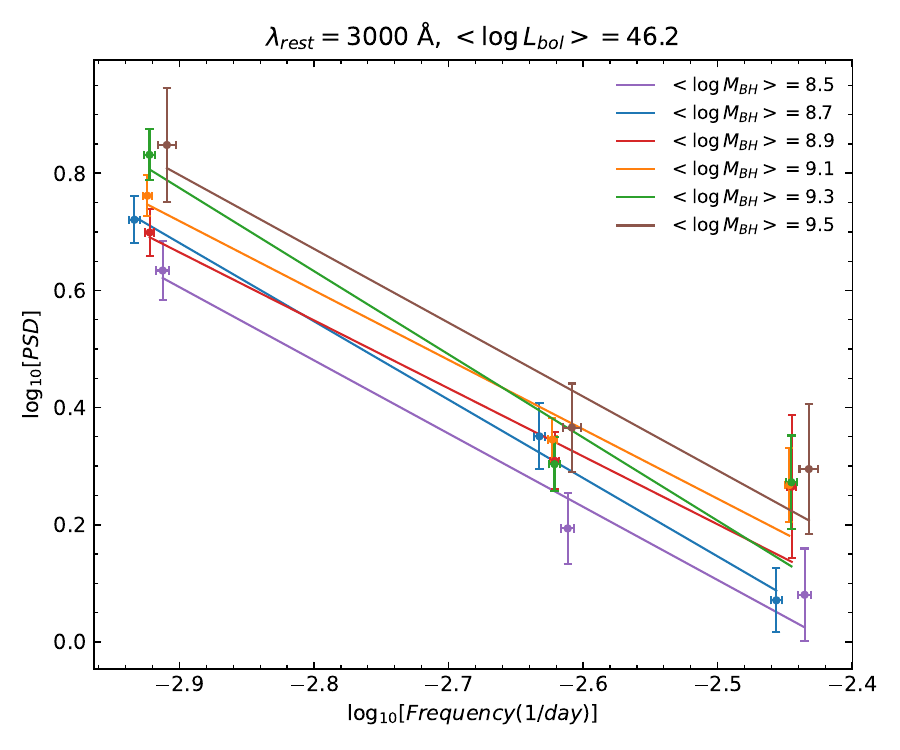}
    \caption{Average PSDs of quasars with 46.1 < \llum \ < 46.3 and all redshifts between 0.5 and 2.5 in different \lbhm \ bins. The legend reports the average mass in each bin of width 0.2.}
    \label{fig:psd_mass}
\end{figure}

Similarly, we considered all quasars in the luminosity bin 46.1 < \llum \ < 46.3 (i.e. the \lum\ bin with the largest number of quasars, see middle panel in Fig.~\ref{fig:quasarpropdistr}) and we computed the mean PSD in narrow BH mass bins. The mass bin size is $\Delta$\lbhm = 0.2, and we used the \lambdarest= 3000\AA\ light curves to compute the PSDs, as before. The resulting PSDs are plotted in Fig.~\ref{fig:psd_mass}. As before with the PSDs in different luminosity bins, the PSDs in the different BH mass bins have similar slopes, but their amplitude appear to depend on \bhm, although not as strongly as the dependence on \lum.

Figure~\ref{fig:psd_mass} shows that the variability amplitude increases with increasing BH mass. This is an unexpected result, as it is difficult to understand how a more massive BH will be more variable than a smaller one. However, since all quasars in the various mass bins have the same \lum, they should have different accretion rates. Assuming that \ledd \ $\propto$ \lum/L$_{\rm Edd} \propto$ \lum/\bhm, then larger BH mass objects in this case should  correspond to objects with smaller accretion rates. Therefore, the apparent trend of \psdamp\ increasing with increasing \bhm\, may in effect be due to the fact that power spectrum amplitude increases with decreasing accretion rate.

In fact, the same trend may be apparent in Fig.~\ref{fig:psd_lum} as well. As luminosity decreases for quasars with the same \bhm, the accretion rate will also decrease. Therefore, the trend of \psdamp\ increase with decreasing \lum\ may be due to the fact that, in reality, \psdamp\ increases with decreasing \ledd. But, when comparing Figs. \ref{fig:psd_lum} and \ref{fig:psd_mass}, it seems that the PSD amplitude depends more strongly on \lum\, than on \bhm. This is despite the fact that \lum\ and \bhm\ vary by the same factor of 10 over the various bins we have considered in Figs. \ref{fig:psd_lum} and \ref{fig:psd_mass}. These black hole and luminosity variations correspond to the same amplitude variations of \ledd. And yet, the \psdamp\ clearly does not vary by the same amount in both cases. This result indicates that \psdamp\ cannot depend only on \ledd, but a joint dependence with \bhm\ remains possible.

\subsection{Modelling the joint dependence on \ledd\ {\it and} \bhm}
\label{sec:psdonbhmandledd}

\subsubsection{PSD amplitude}
\begin{figure}
    \includegraphics[width=1.0\columnwidth]{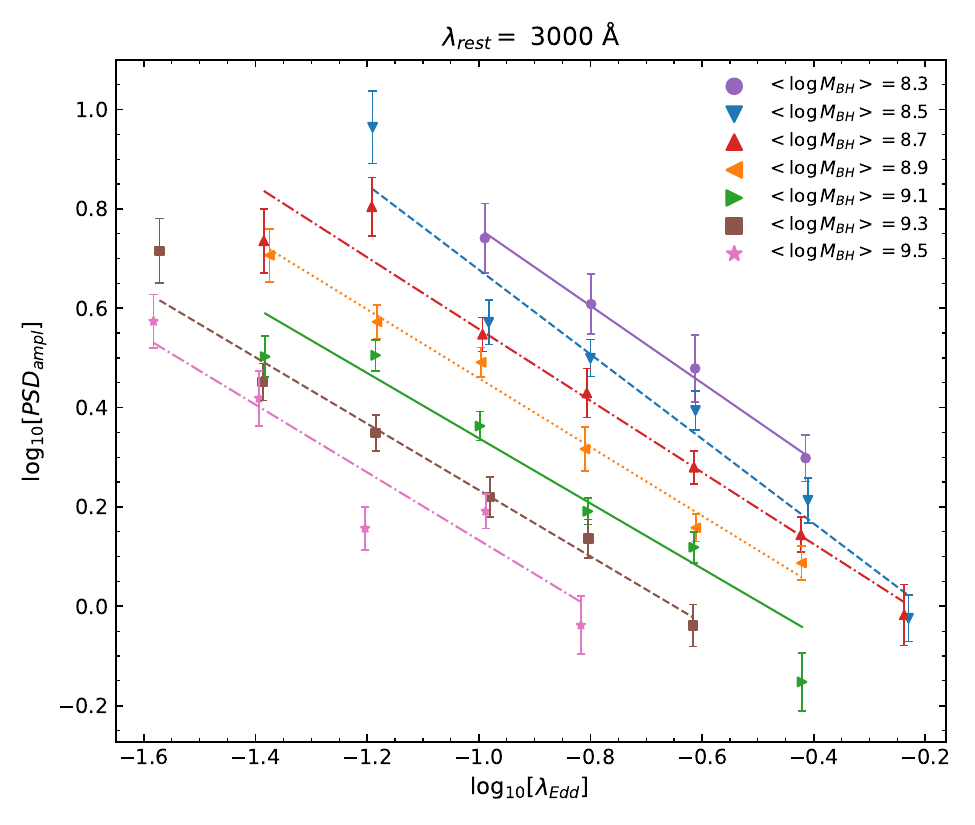}
    \caption{PSD amplitudes for quasars in different \bhm\ bins (points with different symbols and colours) as a function of \ledd. The effect of anti-correlation with luminosity (accretion rate) is still evident, as well as a dependence on \bhm.}
    \label{fig:ledd_trend_all}
\end{figure}

In order to investigate the intrinsic dependence of the PSD amplitude on both \ledd\ and \bhm, we considered quasars in seven different BH mass bins (from 8.2 to 9.6, with $\Delta$\bhm=0.2). We grouped quasars in each BH mass bin in as many luminosity bins as possible (with $\Delta$\lum=0.2), as long as the number of quasars in each bin was at least 50. There are in total 40 different bins. We then computed the mean PSD in each [\bhm--\lum] bin, and we fitted them with the model defined by Eq. \ref{eq:psdmod}. The model fits all PSDs well and all the 40 best-fit \psdamp\ parameters are plotted in Fig. \ref{fig:ledd_trend_all} as a function of log(\ledd). The error on each point is the mean error of all the corresponding 40 best-fit parameters. As we commented above, this error should be more representative of the parameter's uncertainty, than the error of each individual PSD parameter, as computed from the model fit to just three PSD points.

Points with the same symbols and colours, which fall roughly along consecutive diagonal lines, indicate the best-fit \psdamp\ values when we fitted the PSD of quasars which have the same BH mass, but different \lum\, (and hence \ledd). If \psdamp\ depended only on \ledd, then all values should lay along the same diagonal line. This is obviously not the case. The results indicate that \psdamp\ decreases both with increasing \ledd\, as well as with increasing \bhm. 

We fitted the \psdamp\ vs log(\ledd) data in each BH mass bin (i.e. the same colour points in Fig. \ref{fig:ledd_trend_all}) with a line of the form:
\begin{equation}
    {\rm \log_{10}[PSD_{amp}(M_{BH},\lambda_{Edd})]}= {\rm A_{amp}(M_{BH})} + {\rm B_{amp}(M_{BH})}\log\bigg(\frac{\lambda_{\rm Edd}}{0.1}\bigg).
\label{eq:psdampmodel}
\end{equation}
\noindent
The best-fit lines from Eq. \ref{eq:psdampmodel} are shown in Fig. \ref{fig:ledd_trend_all}. They are almost parallel, which suggests that the slope of the dependence of the power spectrum amplitude on \ledd\ is the same for all quasars, irrespective of their BH mass. 

\begin{figure*}
    \includegraphics[width=0.5\linewidth]{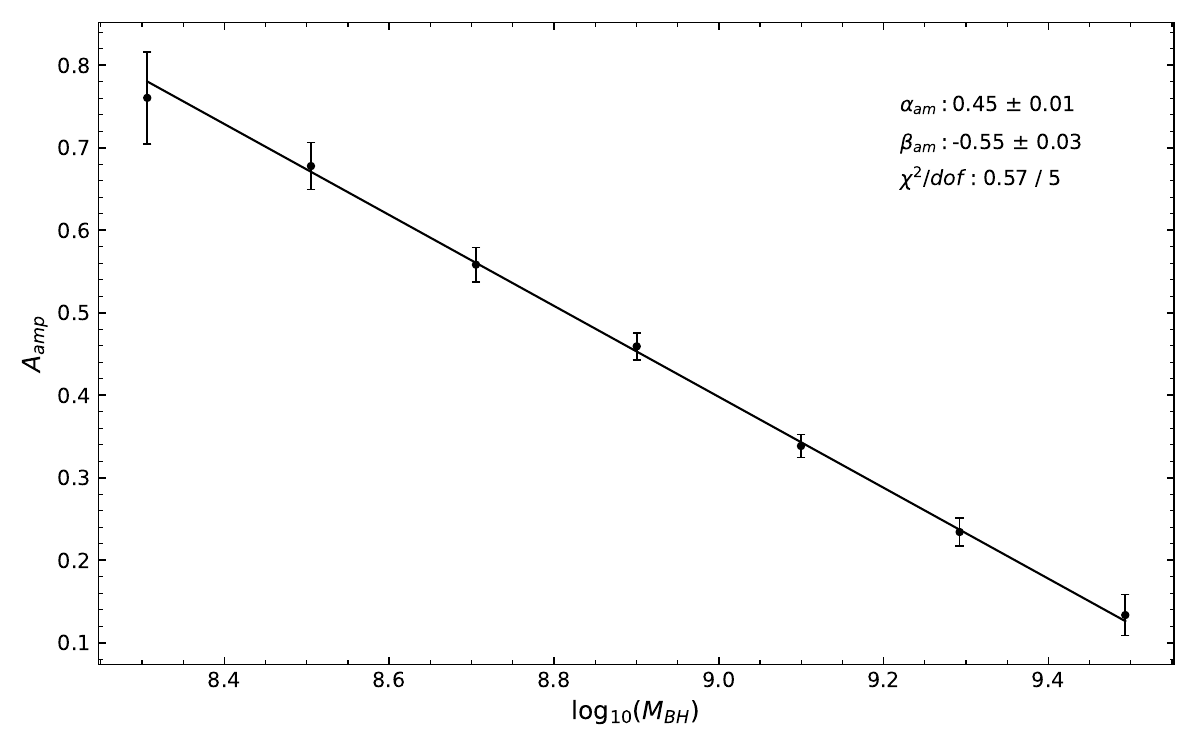}
    \includegraphics[width=0.5\linewidth]{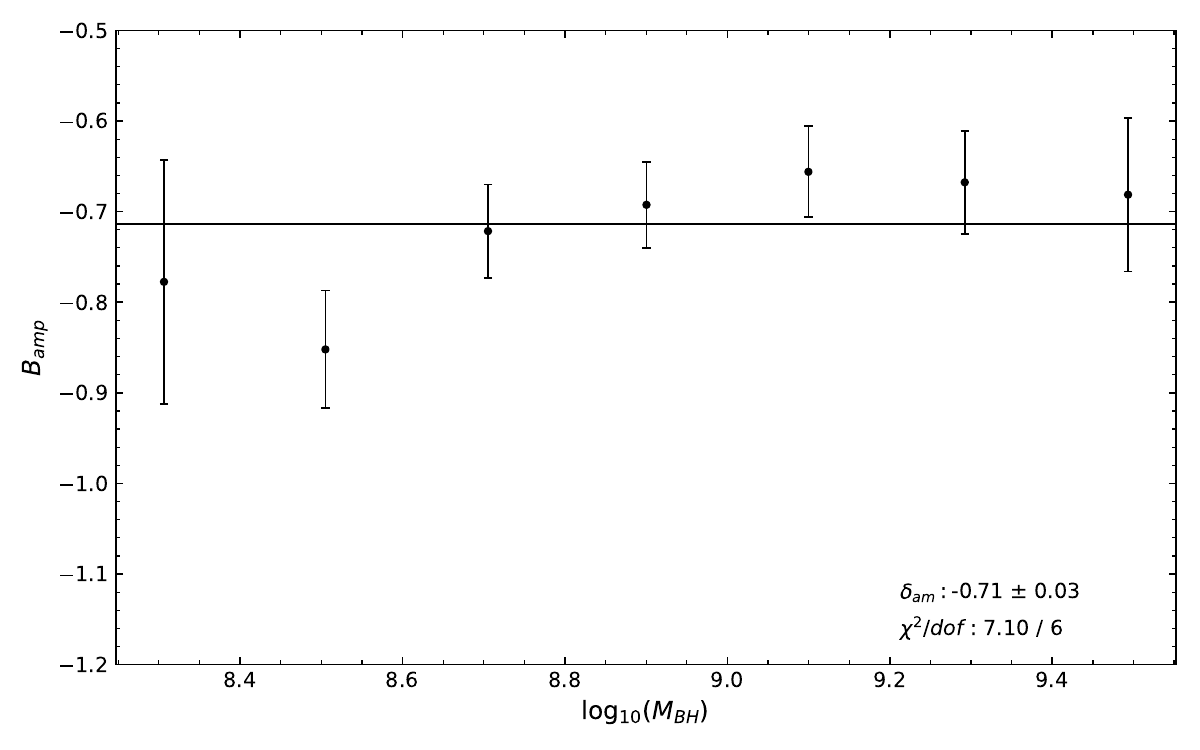}
    \caption{Normalization, i.e.  $A_{amp}$ in Eq.\ref{eq:psdampmodel} (\textit{left panel}), and slope, i.e. $B_{amp}$ in the same equation (\textit{right panel}), plotted as a function of \bhm. The parameters  plotted in these panels determine how the PSD amplitude (plotted in Fig.\,\ref{fig:ampls_all}) depends on BH mass and accretion rate.}
    \label{fig:ampls_all}
\end{figure*}

We plot the best-fit A$\rm _{amp}$ and B$\rm _{amp}$ values as a function of log(\bhm) in Fig.\,~\ref{fig:ampls_all} (left and right panel, respectively). A$_{\rm amp}$ in the left panel is the PSD amplitude at log$(\nu)$= 2.6 (days$^{-1}$), for quasars with \lledd \ = 0.1. Clearly, this amplitude decreases with increasing BH mass. We fitted the A$_{\rm amp}$ vs log(\bhm) data with a line of the form,
\begin{equation}
    {\rm A_{amp}(M_{BH})=\alpha_{am}+\beta_{am}\bigg[\log(M_{BH})-8.9\bigg]}.
    \label{eq:aampeq}
\end{equation}
\noindent This linear model provides an excellent fit to the data, as is shown by the solid line in the left (best-fit statistics reported in the same panel). The best-fit parameters are $\alpha_{am}=0.45\pm0.01$, and $\beta_{am}=-0.55\pm0.03$. On the other hand, B$_{amp}$ does not appear to depend strongly on BH mass, in agreement with what we said above (see text below Eq.\ref{eq:psdampmodel}). A constant line, that is
\begin{equation}
    {\rm B_{amp}(M_{BH})=\delta_{am}},
    \label{eq:bampeq}
\end{equation}
\noindent where $\rm \delta_{am}=-0.71\pm 0.03$ is the mean of the B$_{\rm amp}$ values, provides an acceptable fit to the data (solid line in the right panel in Fig. \ref{fig:ampls_all}; best-fit statistics reported in the same panel).

\subsubsection{PSD slope}
\begin{figure}[t!]
    \includegraphics[width=1.0\columnwidth]{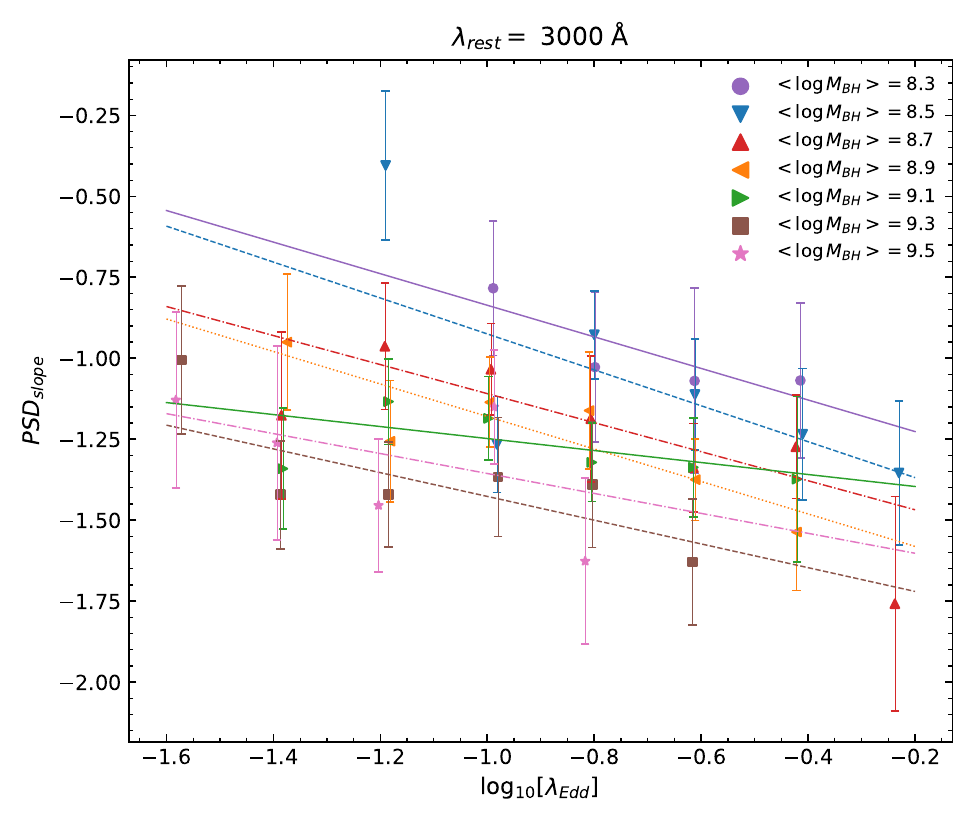}
    \caption{PSD slopes for quasars in different \bhm\ bins (points with different symbols and colours) as a function of \ledd.}
    \label{fig:ledd_trend_all2}
\end{figure}

\begin{figure*}[t!]
    \includegraphics[width=0.5\linewidth]{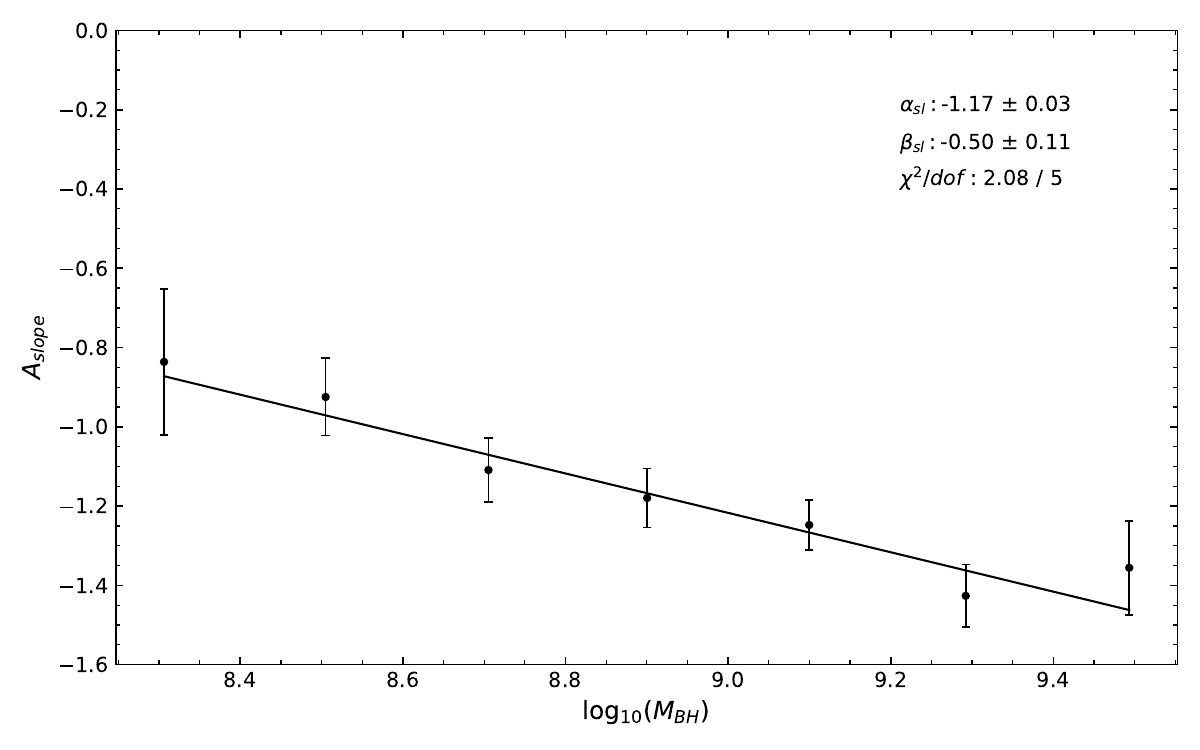}
    \includegraphics[width=0.5\linewidth]{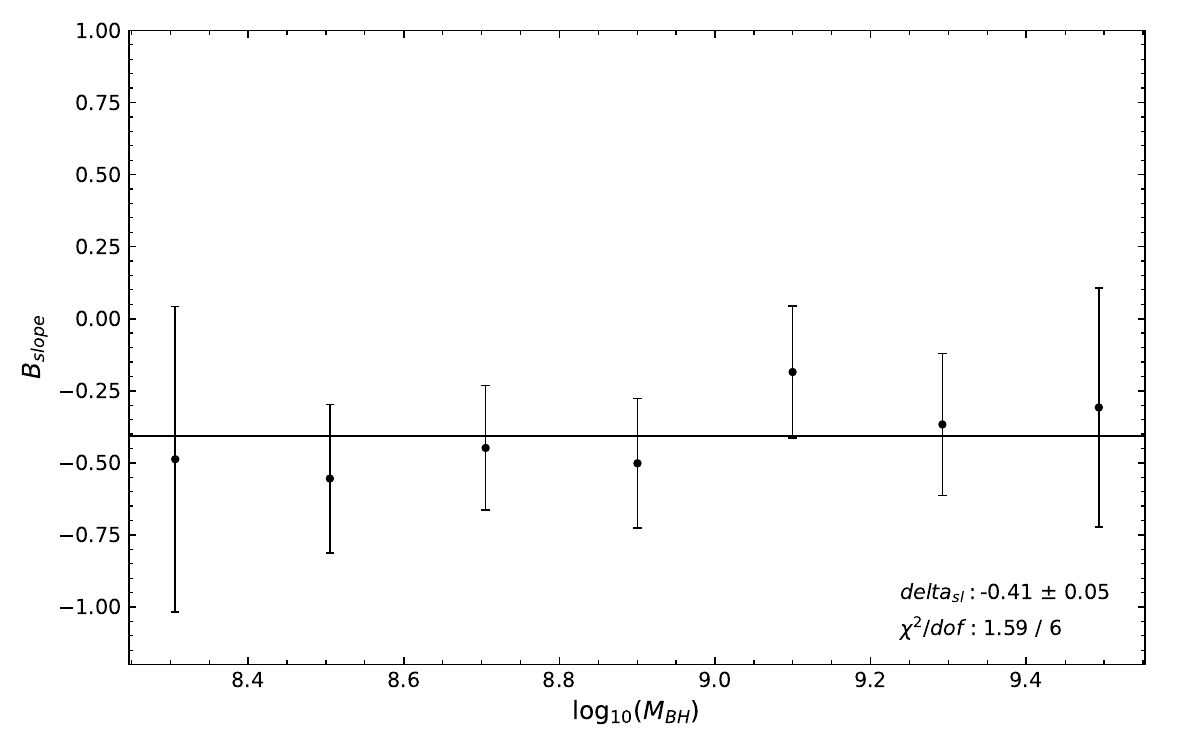}
    \caption{Normalization, i.e.  $A_{slope}$ in Eq. \ref{eq:psdslopemodel} (\textit{left panel}), and slope, i.e. $B_{slope}$ in the same equation (\textit{right panel}), plotted as a function of \bhm. The parameters  plotted in these panels determine how the PSD slope (plotted in Fig.\,\ref{fig:slopes_all}) depends on BH mass and accretion rate.}
    \label{fig:slopes_all}
\end{figure*}

Similar trends, albeit of a smaller amplitude and higher uncertainty, can be found by repeating the same analysis with \psdslope. Figure\,~\ref{fig:ledd_trend_all2} shows the best-fit slopes of the average PSD of quasars in the 40 different bins of \bhm\ and \lum\ defined above (colours and symbols as in Fig. \ref{fig:ledd_trend_all}). As before, if only \ledd\ determined \psdslope, then all points in this plot should be located along the same line, but this does not appear to be the case. We fitted the \psdslope\ vs log(\ledd) data in each BH mass bin (i.e. the same colour points in Fig.\,~\ref{fig:ledd_trend_all2}) with a line of the form: 
\begin{equation}
    {\rm PSD_{slope}(M_{BH},\lambda_{Edd})}= {\rm A_{slope}(M_{BH})} + {\rm B_{slope}(M_{BH})}\log\bigg(\frac{\lambda_{\rm Edd}}{0.1}\bigg).
\label{eq:psdslopemodel}
\end{equation}
\noindent We therefore follow the same approach used for \psdamp\ to analyse A$_{\rm slope}$ and B$_{\rm slope}$ as well. The best-fit A$_{\rm slope}$ and B$_{\rm slope}$ values as function of log(\bhm) are plotted in Fig.\,~\ref{fig:slopes_all} (left and right panel, respectively). A$_{\rm slope}$ in the left panel is the PSD slope for quasars which have different BH mass and a fixed accretion rate \lledd=0.1. Clearly, this slope depends on BH mass: the PSD becomes steeper with increasing BH mass. We fitted the A$_{\rm slope}$ vs log(\bhm) data with a line of the form:
\begin{equation}
    {\rm A_{slope}(M_{BH})=\alpha_{sl}+\beta_{sl}\bigg[\log(M_{BH})-8.9\bigg]}.
    \label{eq:aslopeeq}
\end{equation}
\noindent This line fits the data plotted in the left panel of Fig.\,~\ref{fig:slopes_all} very well (fit statistics are reported on the panel). The best-fit parameters are $\alpha_{sl}=-1.17\pm0.03$, and $\beta_{sl}=-0.5\pm0.1$. On the other hand, B$_{slope}$ does not appear to depend on BH mass: the dependence of the PSD slope on \ledd\ appears to be independent of the quasar BH mass. Indeed, a constant line, that is
\begin{equation}
    {\rm B_{slope}(M_{BH})=\delta_{sl}},
    \label{eq:bslopeeq}
\end{equation}
\noindent where $\rm \delta_{sl}=-0.41\pm 0.05$ is the mean of the B$_{\rm slope}$ values, provides an acceptable fit to the data plotted in the lower panel of Fig. \ref{fig:slopes_all} (best-fit statistics reported in the same panel). 

Taking into account Eqs. \ref{eq:aampeq}, \ref{eq:bampeq}, \ref{eq:aslopeeq}, and \ref{eq:bslopeeq}, then Eqs.\,~\ref{eq:psdampmodel} and \ref{eq:psdslopemodel} can be re-written as follows: 
\begin{equation}
\begin{array}{l}
    {\rm \log_{10}[{PSD_{amp}}(M_{BH},\lambda_{Edd})]= \alpha_{am}} \\[1em]
     \hspace{2em} {\rm + \beta_{am}\bigg[\log\bigg(\frac{M_{BH}}{10^{8.9}}\bigg)\bigg] + \delta_{am}\bigg[\log\bigg(\frac{\lambda_{\rm Edd}}{0.1}\bigg)\bigg]},
\end{array}
\label{eq:psdampmodelfin}
\end{equation}
\noindent and, 
\begin{equation}
    {\rm PSD_{slope}(M_{BH},\lambda_{Edd})= \alpha_{sl} 
    + \beta_{sl}\bigg[\log\bigg(\frac{M_{BH}}{10^{8.9}}\bigg)\bigg] 
    + \delta_{sl}\bigg[\log\bigg(\frac{\lambda_{\rm Edd}}{0.1}\bigg)\bigg]}.
\label{eq:psdslopemodelfin}
\end{equation}
\noindent Equations \ref{eq:psdampmodelfin} and \ref{eq:psdslopemodelfin}, together with the best-fit parameter values listed above, represent our results regarding the power spectrum dependence on BH mass and \ledd, for all quasars in Stripe-82. However, all these results are derived using light curves at distinct filters, as long as their rest-frame wavelength lies within 3000 $\pm$ 300 \AA. We discuss below the dependence of PSD on the rest frame wavelength (i.e. energy) as well.

\subsection{Dependence of PSD on \lambdarest}
\label{sec:wv_depend}
To investigate the PSD dependence on \lambdarest, we repeated the same procedure as in Sect. \ref{sec:psdonbhmandledd}, using light curves of various filters as long as their rest-frame wavelength would be within the \lambdarest\ bins with $|\Delta\lambda|/$\lambdarest = 0.2 indicated by the vertical lines in Fig. \ref{fig:restlambda} (except from, \lambdarest = 3000\AA, of course). So, for each waveband:

\begin{itemize}
    \item we divided all quasars in \bhm\ bins, and each one of them in as many as possible \lum\ bins, as long as there are at least 50 quasars in each [\bhm,\lum] bin;
    \item we computed the mean PSD of all quasars in each bin, and determined \psdamp(\bhm,\lum), and \psdslope(\bhm,\lum);
    \item we created \psdamp\ vs log(\ledd) and \psdslope\ vs log(\ledd) plots, and we fitted the \psdamp\ vs log(\ledd) and the \psdslope\ vs log(\ledd) data (for the same BH mass) using Eqs. \ref{eq:psdampmodel} and \ref{eq:psdslopemodel}; 
    \item we fitted A$_{\rm amp}$, B$_{\rm amp}$, A$_{\rm slope}$ and B$_{\rm slope}$ versus \bhm, and we calculated $\alpha_{am}$, $\beta_{am}$, $\delta_{am}$, and $\alpha_{sl}$, $\beta_{sl}$, $\delta_{sl}$. The respective values for each \lambdarest\ are listed in Table \ref{tab:paramvalues}. 
\end{itemize}

\begin{figure*}
    \includegraphics[width=0.5\linewidth]{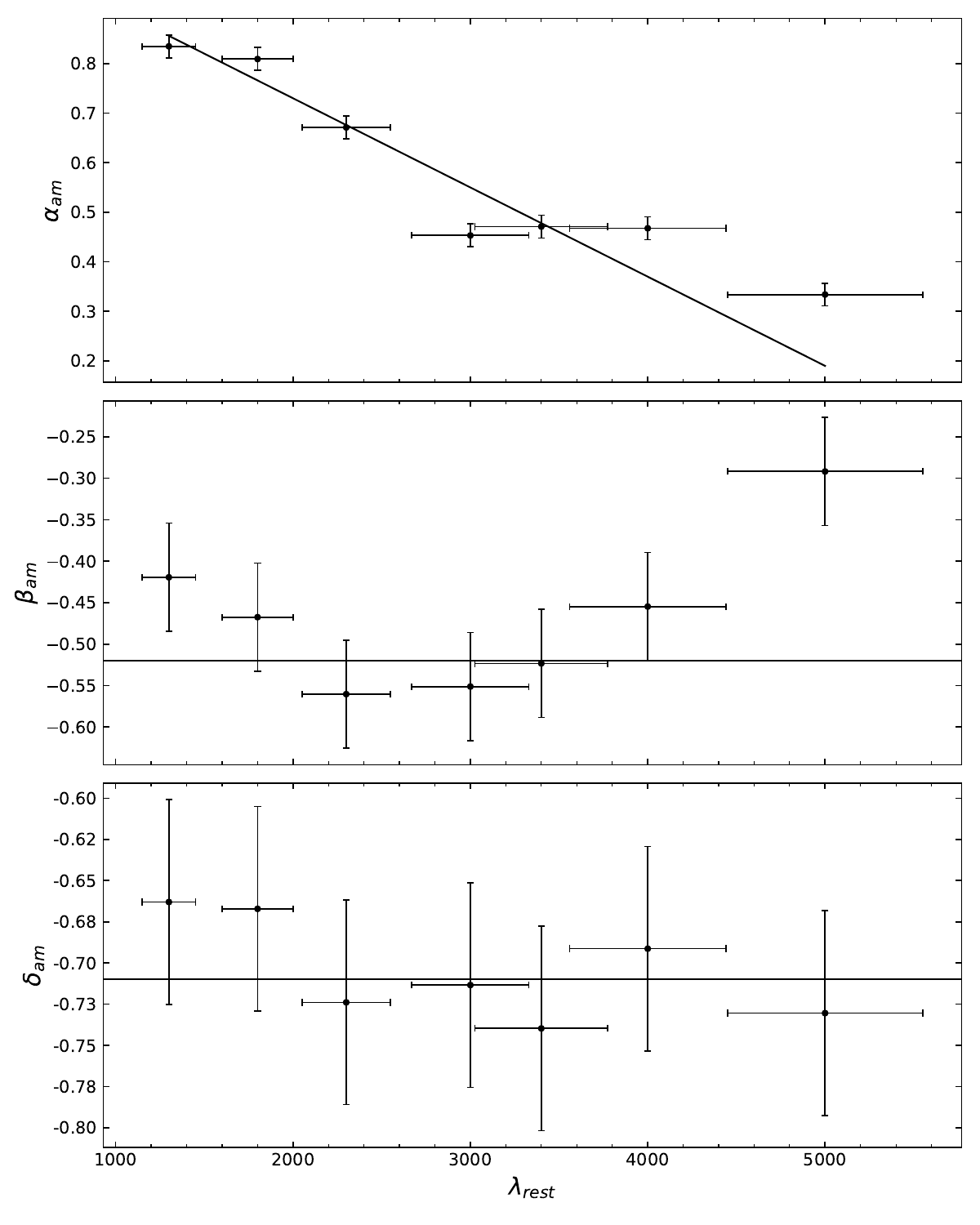}
   \includegraphics[width=0.5\linewidth]{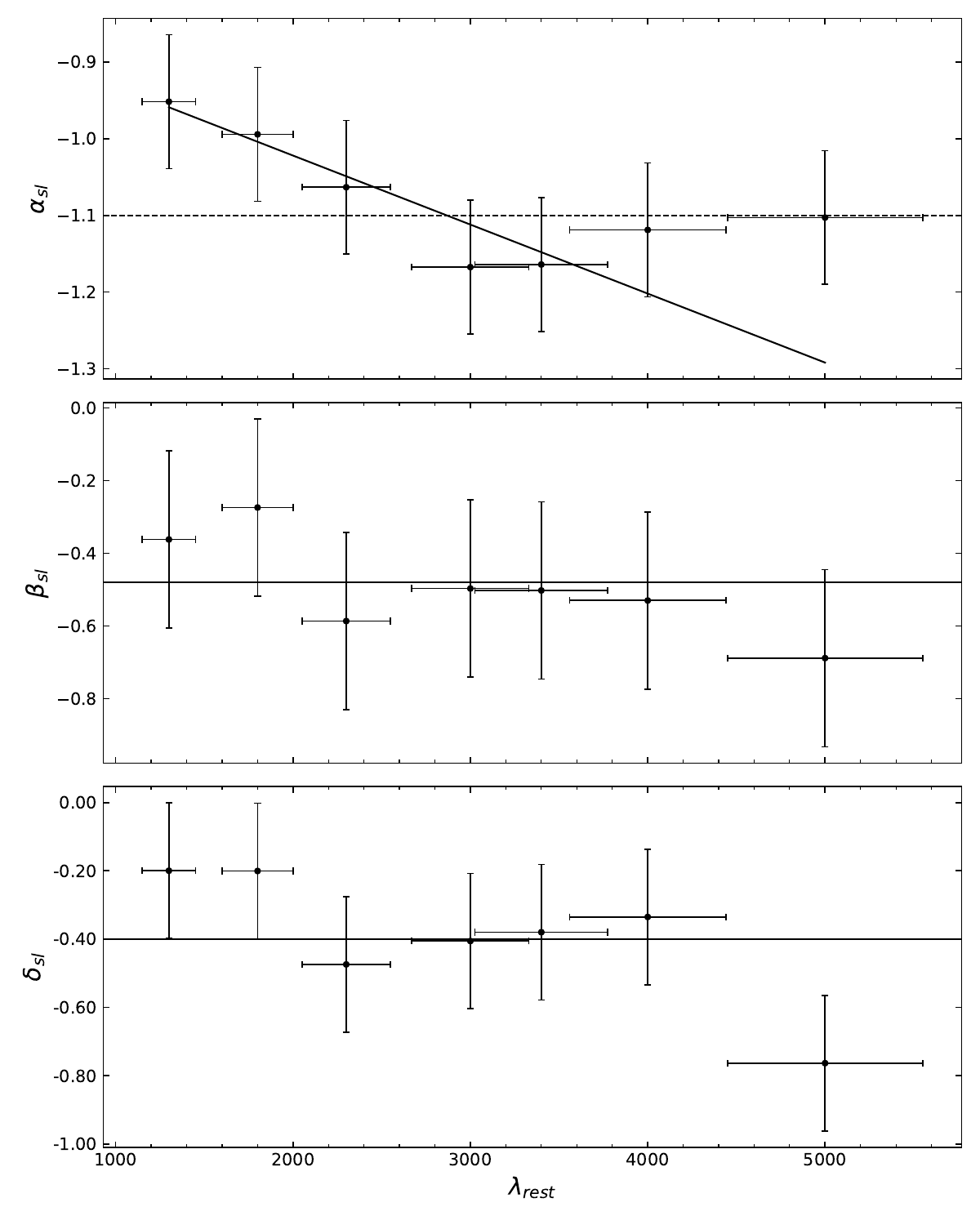}
    \caption{PSD dependence on the rest-frame wavelength. Left panels show the \psdamp\ coefficients $\alpha_{am}$, $\beta_{am}$, and $\delta_{am}$ from Eq.\,\ref{eq:psdampmodelfin} plotted as a function of \lambdarest. Right panels show the same plots for the \psdslope\ coefficients in Eq.\,\ref{eq:psdslopemodelfin}. The solid lines indicate the best-fit lines to the data, the dashed line in the upper-right figure is a weighted mean (see text for details).}
    \label{fig:wv_trends}
\end{figure*}

\noindent Figure \ref{fig:wv_trends} shows a plot of $\alpha_{am}$(\lambdarest), $\beta_{am}$(\lambdarest) and $\delta_{am}$(\lambdarest) versus \lambdarest\ for the \psdamp\ (left panels), as well as $\alpha_{sl}$(\lambdarest), $\beta_{sl}$(\lambdarest) and $\delta_{sl}$(\lambdarest) versus \lambdarest\ for \psdslope\ (right panels). 
Solid lines in each panel represent the best linear fits, obtained by taking into account errors on both axes (where the error bar on the x axis is due to the width of the wavelength bin). In particular, $\beta_{am}$(\lambdarest) and $\delta_{am}$(\lambdarest), as well as $\beta_{sl}$(\lambdarest) and $\delta_{sl}$(\lambdarest) in the middle and lower panels, are acceptably fitted by their weighted means:
\begin{equation}
    \beta_{am}(\lambda_{rest}) = -0.52(\pm 0.01), \
    \delta_{am}(\lambda_{rest}) = -0.71(\pm 0.02), 
\end{equation}
\begin{equation}
    \beta_{sl}(\lambda_{rest}) = -0.48(\pm 0.06), \
    \delta_{sl}(\lambda_{rest}) = -0.40(\pm 0.04).
\end{equation}
This result shows that the dependence of \psdamp\ and \psdslope\ on \ledd\ is the same at each probed wavelengths. 

On the other hand, the plot in the upper-left panel shows that $\alpha_{am}$, defining \psdamp\ (see Eq. \ref{eq:psdampmodelfin}), strongly depends on wavelength. In particular, it decreases with increasing rest-frame wavelength as 
\begin{equation}
{\rm \alpha_{am}(\lambda_{rest}) = -0.55(\pm 0.03) - 1.8(\pm 0.2)\times 10^{-4}}\bigg(\frac{\lambda_{rest}}{3000 \AA}\bigg).
\end{equation}
\noindent
The dependence of the PSD slope, described by $\alpha_{sl}$, is less straightforward. The weighted mean $\Bar{\alpha_{sl}}=-1.10(\pm0.02)$ is not a good fit to the data plotted in the upper-right panel, ($\chi^2=18.2$ for 6 dof, $p_{null}=0.005$). A linear fit returns 
\begin{equation}
{\rm \alpha_{sl}(\lambda_{rest}) = -1.12(\pm 0.02) - 9(\pm 3)\times 10^{-5}}\bigg(\frac{\lambda_{rest}}{3000 \AA}\bigg),
\end{equation}
where the slope is significant just at the $3\sigma$ level while $\chi^2=3.9$ for 5 dof, $p_{null}=0.56$. A line improves the quality of the fit by $\Delta\chi^2=12.6$ for an extra degree of freedom, which is significant according to the F-test (F-statistic $= 18.3$, $p_{null}=0.008$). 

\begin{table*}[ht!]
\caption{Best-fit coefficients of Eqs. \ref{eq:psdampmodelfin} and \ref{eq:psdslopemodelfin} for different \lambdarest defined in Sect. \ref{sec:thesample}.}
\begin{center}
\begin{tabular}{lllllll}
\hline
\hline
\lambdarest & $\alpha_{am}$(\lambdarest)   & $\beta_{am}$(\lambdarest) & $\delta_{am}$(\lambdarest) & $\alpha_{sl}$(\lambdarest) & $\beta_{sl}$(\lambdarest) &  $\delta_{sl}$(\lambdarest)       \\

1300$\pm$130\AA  & 0.83$\pm$0.02 & -0.42$\pm$0.06 & -0.66$\pm$0.09  & -0.95$\pm$0.09  & -0.4$\pm$0.3 & -0.2$\pm$0.2   \\

1800$\pm$180\AA  & 0.809$\pm$0.009 & -0.47$\pm$0.03 & -0.67$\pm$0.04  & -0.99$\pm$0.04  & -0.3$\pm$0.1 & -0.2$\pm$0.2   \\

2300$\pm$230\AA  & 0.671$\pm$0.009 & -0.56$\pm$0.03 & -0.72$\pm$0.05  & -1.06$\pm$0.04  & -0.6$\pm$0.1 & -0.47$\pm$0.08   \\

3000$\pm$300\AA  & 0.453$\pm$0.008 & -0.55$\pm$0.03 & -0.71$\pm$0.03  & -1.17$\pm$0.03  & -0.5$\pm$0.1 & -0.41$\pm$0.05   \\

3400$\pm$340\AA  & 0.471$\pm$0.009 & -0.52$\pm$0.03 & -0.74$\pm$0.04  & -1.16$\pm$0.04  & -0.5$\pm$0.1 & -0.4$\pm$0.2   \\

4000$\pm$400\AA  & 0.47$\pm$0.01 & -0.45$\pm$0.05 & -0.69$\pm$0.04  & -1.12$\pm$0.05  & -0.5$\pm$0.2 & -0.3$\pm$0.1   \\

5000$\pm$500\AA  & 0.33$\pm$0.01 & -0.3$\pm$0.2 & -0.73$\pm$0.15  & -1.1$\pm$0.3  & -0.7$\pm$0.7 & -0.7$\pm$0.6   \\

\hline
\hline
\end{tabular}
\end{center}
\label{tab:paramvalues}
\end{table*}

\section{Discussion and conclusions}
\label{sec:summary}

    \begin{figure*}
    \includegraphics[width=0.5\linewidth]{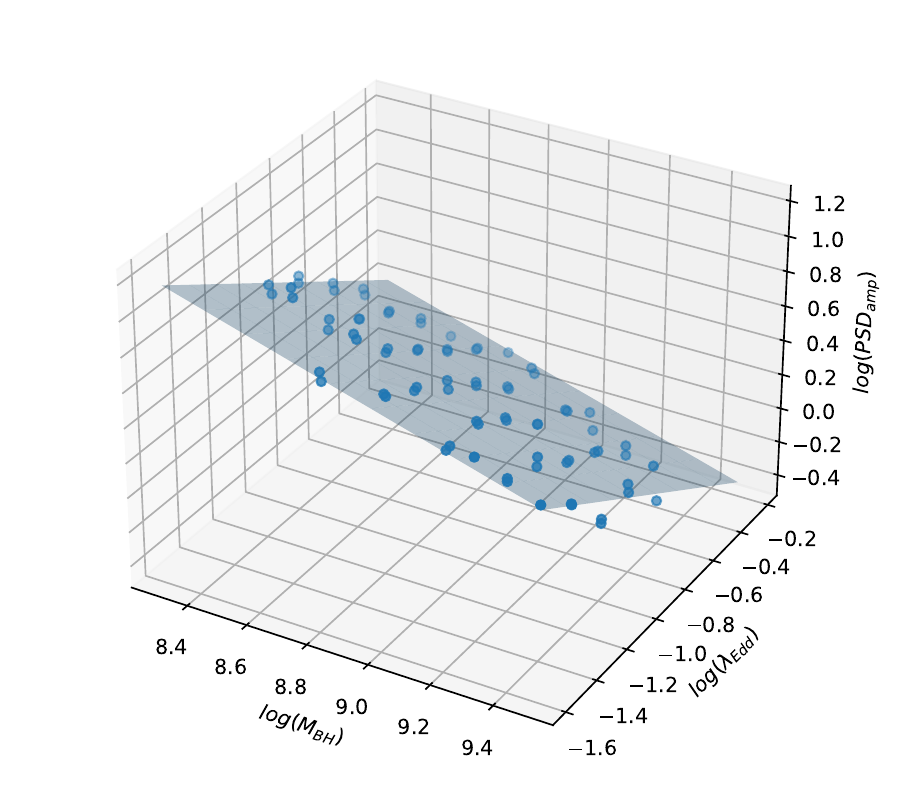}
    \includegraphics[width=0.5\linewidth]{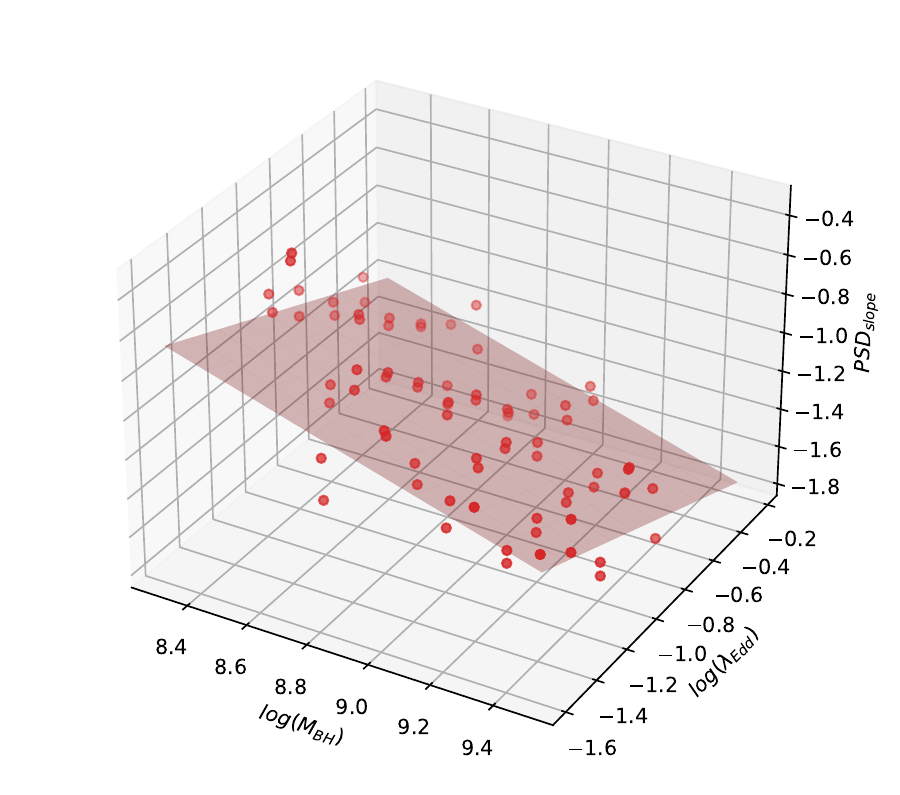}
    \caption{Variability planes for the PSD amplitude and the PSD slope of the average power spectrum of the quasars at \lambdarest=3000\AA, computed using Eqs.\,\ref{eq:psdampmodelfin} and \ref{eq:psdslopemodelfin} (left and right panels, respectively).}
    \label{fig:plane}
    \end{figure*}
    
\begin{figure}
    \includegraphics[width=1.0\columnwidth]
    {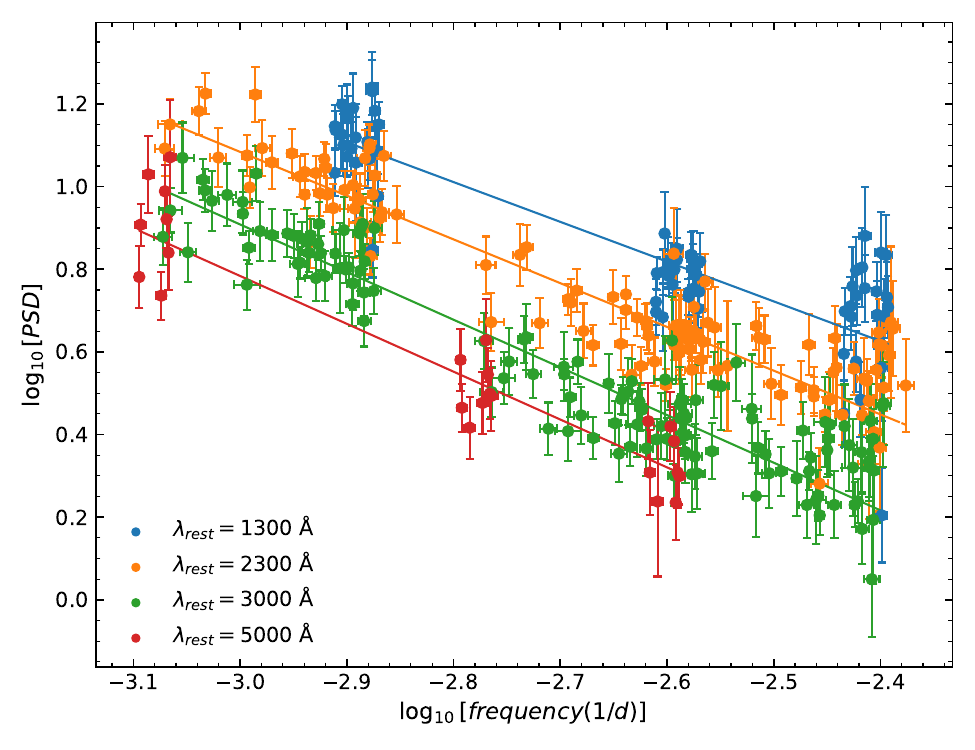}
    \caption{Ensemble PSDs of quasars of all \bhm\ and \ledd, with different \lambdarest\ (indicated with different colours). Power spectra of quasars with various \bhm\ and \ledd\ are normalized to \lbhm\ = 8.9 and \lledd\ = 0.1 using Eq. \ref{eq:psdampmodelfin} and Eq. \ref{eq:psdslopemodelfin}. Solid lines show the best-fit linear models to the data (best-fit slopes are in agreement with the best-fit results plotted in the top panel in Fig.\,\ref{fig:wv_trends}.)}
    \label{fig:psd_combined}
\end{figure}

We studied the continuum UV/optical variability of SDSS Stripe-82 quasars using ensemble power spectral density analysis. The final sample consists of 8042 sources with five nearly simultaneous light curves in \textit{u,g,r,i,z} filters. Each object was observed in at least six yearly seasons, allowing us to sample power spectra over three timescales, in the rest-frame frequency range of $\log{\nu} \sim [-3.2,-2.4]\ day^{-1}$, that is, on timescales of $\sim [250,1500]$ days. The available light curves have been extensively studied in the past but, to the best of our knowledge, this is the first time that they have been used to measure directly the power spectrum of quasars, and investigate how it depends on all major AGN physical parameters, at different UV/optical spectral bands.  Although we estimate the power spectrum of individual objects in just three frequencies, the large number of quasars in the sample,  with parameters distributed over a broad range, allowed us to analyse how variability properties, described by \psdamp\ and \psdslope, depend on quasars physical properties. 

We divided the sample into bins of [\bhm,\lum,z] containing at least 50 sources, and studied the variability properties at seven different rest-frame wavelengths \lambdarest, of width $|\Delta\lambda|/\lambda_{rest}=0.2$ (see Fig. \ref{fig:restlambda}).
\begin{enumerate}
    \item The ensemble power spectrum in all [\bhm,\lum,z] bins is well fitted by a simple power-law of the form:
    \begin{equation}
    {\rm PSD}(\nu) = {\rm PSD}_{amp}(\nu/\nu_0)^{{\rm PSD}_{slope}},
    \label{eq:psdmod_pow}
    \end{equation}
    over the investigated range of frequencies and wavelengths.
    \item The power spectrum shows no significant evidence for a dependence on redshift in the range $0.5 \le$ z $\le 2.5$ at any \bhm, \lum, and \lambdarest\, confirming that quasar variability properties do not evolve with time. Combining all redshifts together, we are able to study the dependencies of the PSD parameters, namely the amplitude and slope, on 
    BH mass ($8.2 \le$ \lbhm\ $\le 9.6$) and accretion rate ($-1.6 \le$ \lledd\ $\le -0.2$).
    \item At each rest-frame wavelength, both \psdamp\ and \psdslope\ show dependencies with \bhm\ and \ledd, independently (see Eq. \ref{eq:psdampmodelfin} and Eq. \ref{eq:psdslopemodelfin}). Such relations define a plane connecting the quasar variability properties (measured through the PSD), with BH mass and accretion rate, as shown in Fig. \ref{fig:plane}. Both the amplitude and the slope are  anti-correlated with mass and accretion rate, i.e. power spectra become steeper and their amplitude decreases for larger black hole masses and higher accreting objects. 

\item We found that the power spectrum also depends on the 
rest frame wavelength. This is shown in Fig. \ref{fig:psd_combined} where we plot the ensemble PSD of quasars of all BH masses and accretion rates, but at different \lambdarest\ (PSDs for the various \bhm\ and \ledd\ were normalized to \lbhm=8.9 and \lledd=0.1 using Eq. \ref{eq:psdampmodelfin} and Eq. \ref{eq:psdslopemodelfin}). Clearly, the PSD amplitude decreases significantly with increasing wavelength. We find some indication that \psdslope\ may also depend on \lambdarest, with power spectra becoming steeper at longer wavelengths. However, the significance for this effect is not as strong as for the other relation, and any PSD slope variations should be of relatively small amplitude (see upper panels of Fig. \ref{fig:wv_trends}).

\end{enumerate}

Our final results regarding \psdamp\ and \psdslope\ can be summarized by the following equations:
\begin{equation}
\begin{array}{l}
    \rm \log_{10}{[PSD_{amp}(M_{BH},\lambda_{Edd}, \lambda_{rest})]} = -0.55(\pm 0.03)   \\[1em] -1.8(\pm 0.2)\times10^{-4}\bigg(\frac{\lambda_{rest}}{3000\AA}\bigg) - 0.52(\pm 0.01)\bigg[\log_{10}\bigg{(\frac{M_{BH}}{10^{8.9}}\bigg)}\bigg]  \\ - 0.71(\pm 0.02)\bigg[\log_{10}\bigg{(\frac{\lambda_{Edd}}{0.1}\bigg)}\bigg],
\end{array}
\label{eq:psdamp_summary}
\end{equation}
and

\begin{equation}
\begin{array}{l}
    \rm PSD_{slope}(M_{BH},\lambda_{Edd}, \lambda_{rest}) = -1.12(\pm 0.02)  \\ -9(\pm 3)\times10^{-5}\bigg(\frac{\lambda_{rest}}{3000\AA}\bigg) - 0.48(\pm 0.06)\bigg[\log_{10}\bigg{(\frac{M_{BH}}{10^{8.9}}\bigg)}\bigg]  \\ - 0.40(\pm 0.04)\bigg[\log_{10}\bigg{(\frac{\lambda_{Edd}}{0.1}\bigg)}\bigg].
\end{array}
\label{eq:psdslope_summary}
\end{equation}

\noindent
In using these results to constrain physical models of quasar variability, it must be stressed that our relations characterize the quasar PSD properties on timescales of 2-6 years (observer frame), and were derived studying quasars with luminosity ($45.3 \le$ \llum\ $\le 46.5$), BH mass ($8.2 \le$ \lbhm\ $\le 9.6$) and accretion rate ($-1.6 \le$ \lledd\ $\le -0.2$).

\subsection{Comparison with previous studies}
\label{sec:comparison}

We analyse the same sample of SDSS quasars as \cite{MacLeod10}. However, while we used a model-independent approach and measured power spectra directly from the data, they assumed a DRW model and found its best-fit parameters, namely the decorrelation time $\tau$ and the long-term rms variability amplitude {\sl SF}$_{\infty}$. In general, our results are in agreement with \cite{MacLeod10} for what concerns the trends of the variability parameters (the PSD amplitude and slope, in our case) on AGN luminosity, BH mass, accretion rate and rest-frame wavelength. Like \cite{MacLeod10}, we also find that these parameters do not depend on redshift. 

\begin{figure}
    \includegraphics[width=1.0\columnwidth]
    {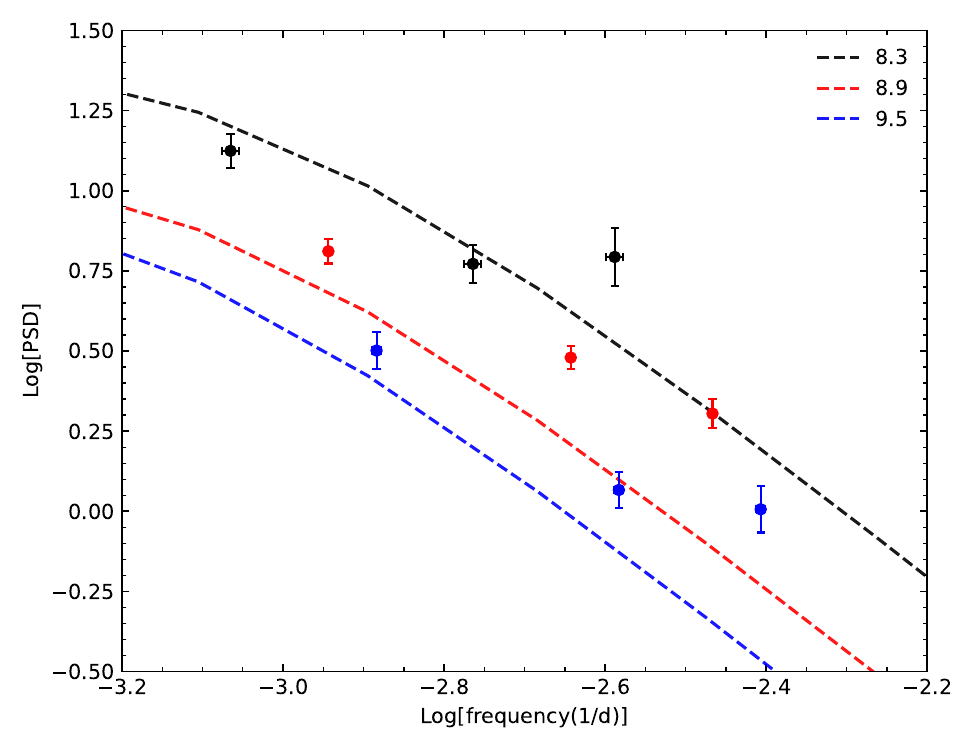}
    \caption{Power spectra of quasars  at 3000\AA\ with \lledd\ = 0.1 and three different mass bins: \lbhm\ = 8.3, 8.9, and 9.5 (different colours). Points show PSD values measured in this work, while dashed lines are the best-fit DRW models from \cite{MacLeod10}.}
    \label{fig:macleod}
\end{figure}

On the other hand, we find PSD slopes around $-0.8/-1.2$ (see left panel of Fig. \ref{fig:slopes_all}), depending on BH mass, while the DRW model implies that the PSD slope is $-2$, irrespective of the BH mass, above the damping frequency. This discrepancy could imply a problem with the DRW assumption, as other works also found flatter slopes from ensemble studies (e.g. \citealt{VandenBerk+2004}). However, as we have stressed before, our results are based on the PSD analysis in a particular frequency range. It is possible that this frequency range is close to the range where the DRW power spectrum bends from a slope of $-2$ to zero, which could explain the flatter PSD slopes. In order to test this possibility, we used the scaling relations provided by \citet[][their Eq.7 and Table 1]{MacLeod10} to derive the DRW parameters for quasars with \lledd\ = 0.1 and \lbhm\ = 8.3, 8.9, and 9.5, at 3000\AA. We then determined the PSD resulting from the DRW model for these parameters, following Eq. 2 of \cite{Kozlowski2010}: 
\begin{equation}
    P(\nu)=\frac{2\hat{\sigma}^2\tau^2}{1+(2\pi\tau\nu)^2},
\label{eq:psd_drw}
\end{equation}
\noindent
where $\hat{\sigma}=\mbox{\sl SF}_{\infty}/\sqrt{\tau}$. Figure \ref{fig:macleod} shows the resulting DRW models (dashed lines) and the PSDs we have measured for quasars with the same BH mass and accretion rate (points and lines with the same colours correspond to models and measured PSDs of quasars with the same parameters). 

Model predictions do not seem to be in good agreement with our power spectrum measurements as both the relative amplitudes and the slopes are different. The flatter slopes might be partly explained considering aliasing effects on the power spectra; however in order to be consistent with DRW models we need to have the DRW break exactly in the sampled frequency range (see Appendix \ref{app_B}). In fact, even \cite{MacLeod10} were not able to distinguish between a spectral index of --2 or --1 with the limited Stripe-82 sampling, albeit they assumed a DRW to derive scaling relations. Furthermore, the damping timescales predicted by \cite{MacLeod10} are likely underestimated due to the finite length of light curves (see discussion in \citealt{kozlowski+17} and \citealt{Suberlak+21}). 
If this was the case, the discrepancy between DRW and our results would increase, as model slopes would be closer to --2 and aliasing effects on our power spectra would be minimal. We therefore conclude that DRW are not fully consistent with our results, although putting definitive constraints on the intrinsic PSD shape (i.e. whether and where it is bending) with Stripe-82 data is difficult without making further assumptions.

\begin{figure}
    \includegraphics[width=1.0\columnwidth]
    {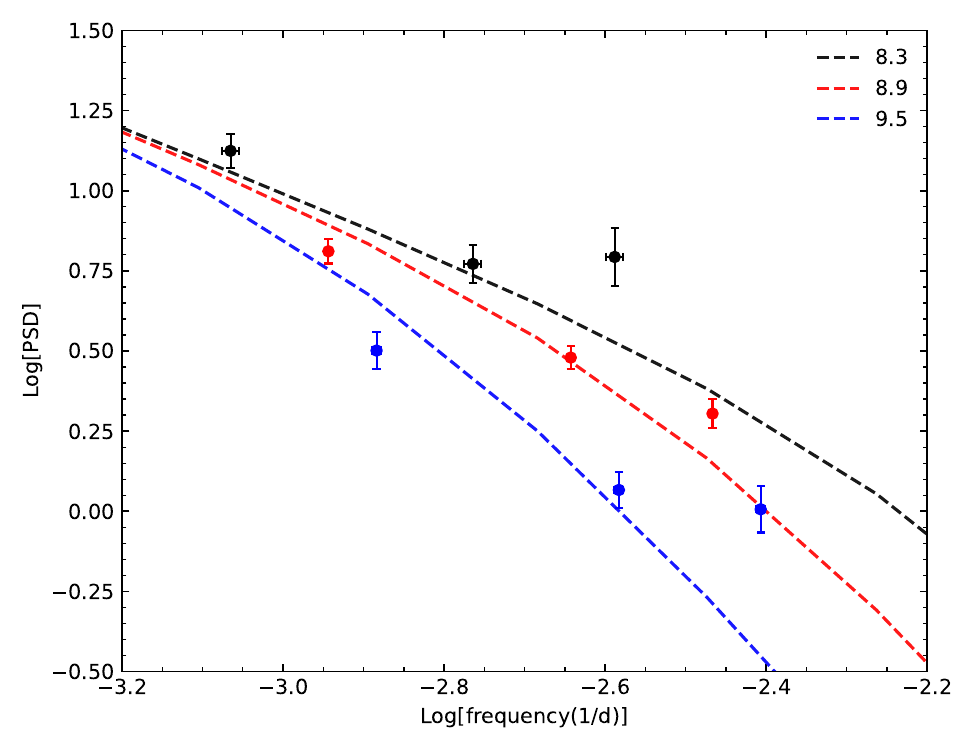}
    \caption{Power spectra of quasars  at 3000\AA\ with \lledd\ = 0.1 and three different mass bins: \lbhm\ = 8.3, 8.9, and 9.5 (different colours). Points show PSD values measured in this work, while dashed lines are the best-fit models from \cite{Ar2023}, at 2900\AA.}
    \label{fig:arevaloo}
\end{figure}

Recently, \cite{Ar2023} modelled the quasar variability using data from the ZTF archive, with a range of physical properties similar to those analysed in our work. They found that the PSD is well fitted by a broken power law, where both the amplitude and the break frequency depend on BH mass and accretion rate. The best-fit model predicts power indices of --3 (at high frequencies) and --1 (at low frequencies), while the DRW model resulted in a significantly worse fit. Figure \ref{fig:arevaloo} shows a comparison between the PSD of the SDSS quasars from our analysis and their model predictions. 
Data points and dashed lines are as defined in Fig. \ref{fig:macleod}. The overall absolute normalization of the \cite{Ar2023} models has been arbitrarily shifted to broadly match our PSD estimates, while the relative differences are constrained by the scaling relations derived in \cite{Ar2023}. We also point out that these relations are defined at $\lambda_{rest}=2900\AA$, while our PSD estimates are at $\lambda_{rest}=3000\AA$. 

The \cite{Ar2023} model predictions also do not seem to reproduce well our results, albeit the discrepancy seems to be smaller than with the DRW models from \cite{MacLeod10}. Our flatter slopes seem to suggest a higher frequency break although, again, the limited frequency range explored in our analysis does not allow to robustly constraint the position of the break. A more thorough analysis on both longer and shorter timescales will be subject to a future work.

\subsection{Addressing the meaning of the scaling relations}
\label{sec:meaning}
The scaling relations presented in this work describe the dependence between the main physical parameters of quasars and their variability properties, as determined by a model independent power spectrum analysis. As we argued above, such relations are important, as they should allow us to test different theoretical models. Testing specific models is beyond the scope of the present work. However, we can use the scaling relations and try to understand, in broad terms, possible physical reasons behind them. 

\subsubsection{The dependence of PSD amplitude and slope on BH mass}
The anti-correlation of \psdamp\ and \psdslope\ with BH mass could be due to the fact that the power spectrum is the same in all AGN, but the intrinsic timescales are proportional to \bhm. This may be natural, as a larger BH mass implies a larger size for the accretion disc itself (the gravitational radius scales proportionally to \bhm). After all, the accretion disc characteristic timescales are proportional to \bhm. In this case, it is possible that the relations we found may be due to the fact that the observed power spectra sample different intrinsic physical timescales in each source. Similar considerations have also been proposed in other recent works using different data, such as \cite{Ar2023} and \cite{Tang2023}.

\begin{figure}
    \includegraphics[width=1.0\columnwidth]{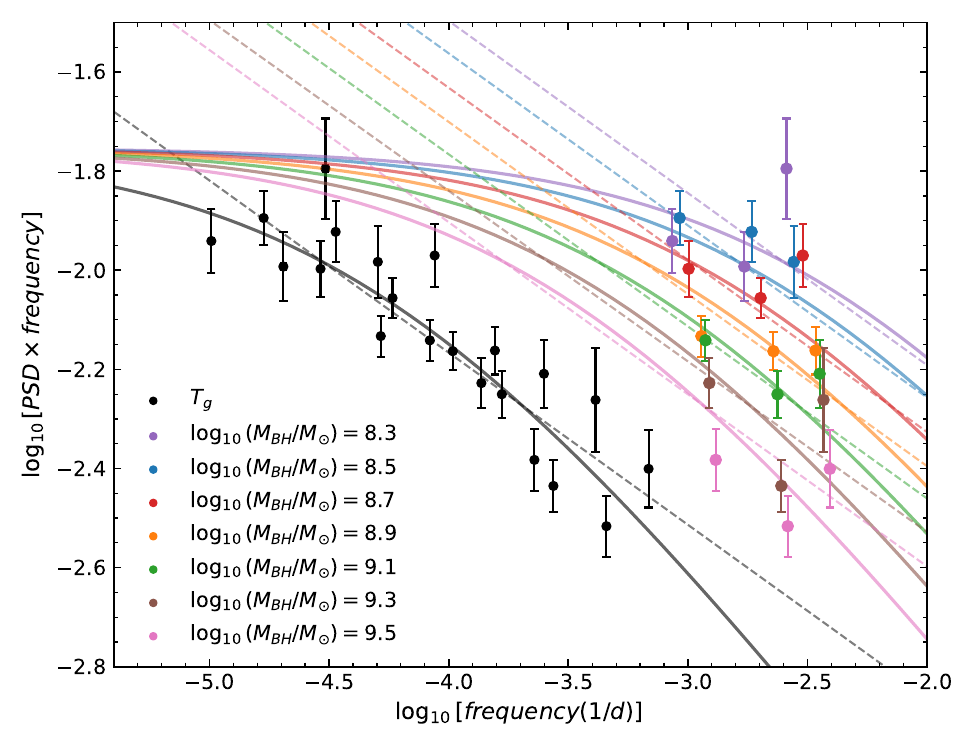}
  \caption{Ensemble PSDs of quasars with \lambdarest\ = 3000\AA, \lledd\ $= -1.0$ and various  \bhm\ (points with different colours on the plot). Black points indicate the universal PSD of all the quasars, computed when light curves are normalized with the gravitational timescale, $T_g$. Black dashed and solid lines show the best-fit single and bending power law models to the black points, respectively. Coloured lines show the same best-fit models re-scaled in frequency by $1/T_g$, for each of the BH mass listed in the bottom left part of the plot (see Sect. \ref{sec:meaning} for details).}
    \label{fig:psd_scalings3}
\end{figure}

In order to test this idea, we transformed the observed light curve of each object to an intrinsic light curve by dividing the observed times by the light travel time of the gravitational radius of the central BH, that is
\begin{equation}
    T_g = \frac{GM_{BH}}{c^3},
\label{eq:tg}
\end{equation}
\noindent where $G$ is the universal constant of gravitation, 
and $c$ is the speed of light. As our power spectra are computed using the periodogram defined in Eq. \ref{periodogram}, this normalization implies that both the sampled frequencies and the power spectrum amplitude are  respectively multiplied and divided by $T_g$. 
We applied this normalization to seven groups of quasars with  8.2 < \lbhm\ < 9.5 and \lledd\ = -1.0, using light curves at 3000\AA\ rest-frame. 
The result is shown in Fig. \ref{fig:psd_scalings3}. 

Points with different colours in this figure show the original (i.e. non-scaled) power spectra of the different BH mass groups (note that the y-axis in this figure shows $\log_{10}{[PSD(\nu) \times \nu]}$). Black points show the scaled power spectra, as we explained above.  The alignment of all the PSD estimates after the normalization with $T_g$ is quite good. The fact that all the points become almost aligned after the PSD normalization with $T_g$ indicates that the assumption of a power spectrum which has the same shape in all quasars, irrespective of their BH mass, while timescales depend linearly on \bhm\ is reasonable.

The dashed black line in Fig.\,\ref{fig:psd_scalings3} shows the best linear fit to the black points.  We also fitted the power spectrum with a  bending power law \cite[e.g.][]{McHardy04}:
\begin{equation}
    P(\nu) = N\nu^{-1}\bigg[1+\bigg(\frac{\nu}{\nu_b}\bigg)^{\alpha-1}\bigg]^{-1},
\label{eq:brokepsd}
\end{equation}
\noindent as shown by the solid black line. This is a model that fits well the X--ray PSDs of nearby Seyferts. It corresponds to a power law of slope --1 at low frequencies, which steepens to a
slope of $\alpha$ at frequencies higher than $\nu_b$. 
The best-fit slope of the linear fit is $-1.34(\pm0.03)$, while the best-fit high frequency slope of the bending power law model is $-1.6(\pm0.1)$, and  $\log_{10}(\nu_b)=-2.90(\pm0.01)$. The high frequency slope is rather flat, and it is for this reason that both the linear and the bending power-law models fit the data equally well. Statistically speaking, neither of the two fits are acceptable ($\chi^2=46.5$/19 dof and $\chi^2=40.6$/18 dof for the linear and the bending power-law fits, respectively). However, as the figure shows, both models fit the normalized PSD rather well,  with no systematic trends in the best-fit residuals. 

The solid and dashed coloured lines in Fig. \ref{fig:psd_scalings3} show the best-fit power law and bending power fits to the normalized PSD (i.e. the black solid and dashed lines) scaled back to the PSD of the various mass groups plotted with the same colours. To re-scale the PSD models we have kept the best-fit parameters the same, and we re-scale the frequencies by 1/$T_g$, that is
\begin{equation}
    \nu_{M_{BH}}= \nu_{norm} \frac{1}{GM_{BH}/c^3},
\end{equation}
\noindent where $\nu_{norm}$ are the frequencies of the universal quasar PSD (shown with the black points in Fig. \ref{fig:psd_scalings3}), and $\nu_{M_{BH}}$ the frequencies of the PSD of quasar of mass \bhm. The re-scaled PSDs agree well with the observed PSDs, as expected. Again, this supports the existence of a universal PSD shape for all quasars, where frequencies scale with BH mass, while normalization and slope(s) are fixed (at any given wavelength and accretion rate).

\subsubsection{The dependence of PSD on \ledd, and \lambdarest}
Even if the hypothesis of a universal PSD which scales with BH mass in frequency could explain the dependence of the power spectrum on \bhm, it is difficult to understand the dependence on accretion rate. As the accretion rate increases, so does the surface area of the disc that emits at a given wavelength.  If the UV/optical variability is due to variations that take place in the inner disc only (i.e. at distances within a given radius same in all objects in $R_g$ units), then it is natural to expect that the variability amplitude should decrease as the disc emitting area increases
(i.e. for larger accretion rates). 
At the same time, the power spectrum amplitude is expected to decrease with increasing \ledd\, as observed, in the case of X--ray reverberating discs. This is because the amplitude of the disc transfer function decreases with increasing accretion rate, at all wavelengths (see e.g. Fig. 10 in \citealt{Panagiotou22}). 

The dependence on \lambdarest\ may be simpler to understand. In the case of X-ray reverberating discs, the amplitude of the disc transfer function decreases with increasing wavelengths, which implies that the PSD amplitude should also decrease (see e.g. all the figures in Appendix B of \citealt{Panagiotou22}). In general, the disc-emitting area increases with increasing wavelength. Again, as we discussed above, it may be natural to expect that the variability amplitude would decrease as the disc emitting area increases (i.e. for longer wavelengths). 

The discussion above does not propose a specific theoretical model for the observed UV/optical variability of quasars. It is meant to highlight a few of our results and demonstrate some potential, general constraints placed on various theoretical ideas that have been proposed in the past. We believe that, as long as the observed variations are associated with physical mechanisms that operate in the innermost region of quasars, then our results, i.e. the dependence of the PSD on BH mass, accretion rate, and rest frame energy, can constrain, strongly, any physical model for the UV/optical variability of quasars.

\begin{acknowledgements}
We thank the anonymous referee for the useful comments and suggestions. MP, VP and DD acknowledge the financial contribution from PRIN-MIUR 2022 and from the Timedomes grant within the ``INAF 2023 Finanziamento della Ricerca Fondamentale''. VP also acknowledges Erasmus+ learning mobility founds 2022/2023 and the Insitute of Astronomy at FORTH, Greece, for hospitality. DD also acknowledges PON R\&I 2021, CUP E65F21002880003.
FEB acknowledges support from ANID-Chile BASAL CATA FB210003, FONDECYT Regular 1200495 and 1241005, and Millennium Science Initiative Program  – ICN12\_009. This work has been partially supported by ICSC – Centro Nazionale di Ricerca in High Performance Computing, Big Data and Quantum Computing, funded by European Union – NextGenerationEU.
\newline
The research has made use of the following \texttt{Python} software packages: \texttt{Matplotlib} (\citealt{Hunter2007}), \texttt{Pandas} (\citealt{McKinney2010}), \texttt{NumPy} (\citealt{vanderwalt2011}), \texttt{SciPy} (\citealt{Virtanen2020}).

\end{acknowledgements}

\bibliography{references.bib}
\bibliographystyle{aa.bst}

\begin{appendix} 

\section{Measurement of Poisson noise in the PSD}
\label{app_A}
We considered a sample of $\sim5\times10^5$ non-variable stars in Stripe-82 (\citealt{ivezic2007}) to model the experimental Poisson noise in the quasars PSD. For each one of the five bands \textit{ugriz}, we divided the sample of stars in bins of magnitude of width 0.2 and measured their PSD with the same prescription that we used to measure the quasars PSDs (i.e. we binned the light curves, computed the PSD at three frequencies for each star and then we constructed the ensemble PSD for the selected magnitude bin - see Sect. \ref{sec:psdestimation} for details). 

\begin{figure}
    \includegraphics[width=1\columnwidth]{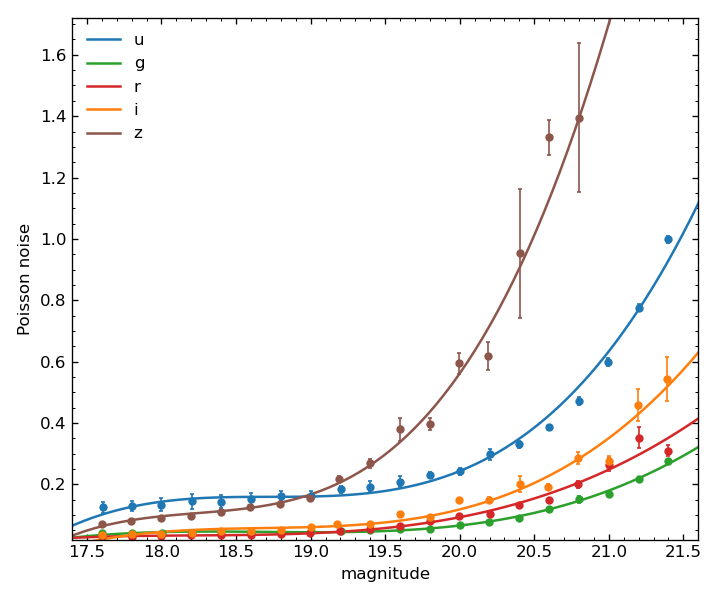}
    \caption{White noise contribution as a function of magnitude for non-variable stars in the five SDSS filters. Continuous lines show third-order polynomial best-fits to the data.}
    \label{fig:stars_psd}
\end{figure}

Since we selected non-variable stars, we expect the ensemble PSD to be flat in all magnitude bins and spectral bands. We calculated the mean of the ensemble PSDs in each bin and we assume that this mean value is an accurate estimate of the Poisson noise contribution in each spectral band and magnitude range. Figure \ref{fig:stars_psd} shows the mean ensemble PSD in each magnitude bin plotted as a function of magnitude for each filter. The solid lines show the best-fit of a third-order polynomial function to the data. As the Poisson noise should depend on the object's flux only, we used the best-fit lines to predict the Poisson noise level of each quasar in our sample, according to its magnitude in each filter. Such a noise value is then subtracted from the PSD estimates obtained with Eq. \ref{periodogram}, so that:
\begin{equation}
    PSD_{corrected} = PSD_{original} - noise.
\end{equation}

\section{PSD of simulated light curves}
\label{app_B}
As we mentioned in Sect. \ref{sec:psdestimation}, the periodogram is the usual estimator of the intrinsic PSD of a variable process. Its statistical properties are known very well (e.g. \citealt{Priestley1981}) and they have been studied extensively even in the case of red-noise like processes which are quite common in astronomy (e.g. \citealt{1993MNRAS.261..612P} and references therein). We use light curves with six points to compute the periodogram in three frequencies, for each quasar in the sample. Nevertheless, irrespective of the small number of points in the light curves, the periodogram statistical properties are only marginally affected.

To demonstrate this, we simulated stochastic light curves with a cadence of 1 day and a length of 20 years, assuming a DRW model with damping timescales in the range $[10^2-10^5]$ days (i.e. we went from a broken power-law, to a single power law with a slope of --2). The procedure for simulating DRW light curves is described in \cite{MacLeod10} and \cite{Kovacevic2021}. For each fixed DRW $\tau$ and $\sigma$, we created 50 simulated light curves and we applied the same sampling pattern as that of the observed SDSS Stripe-82 light curves. We then processed the simulated data with the same procedure applied to real data (i.e. we binned the time series in yearly seasons) and we computed the periodograms at three frequencies with the resulting light curves which consist of six points.

\begin{figure}
    \includegraphics[width=1\columnwidth]{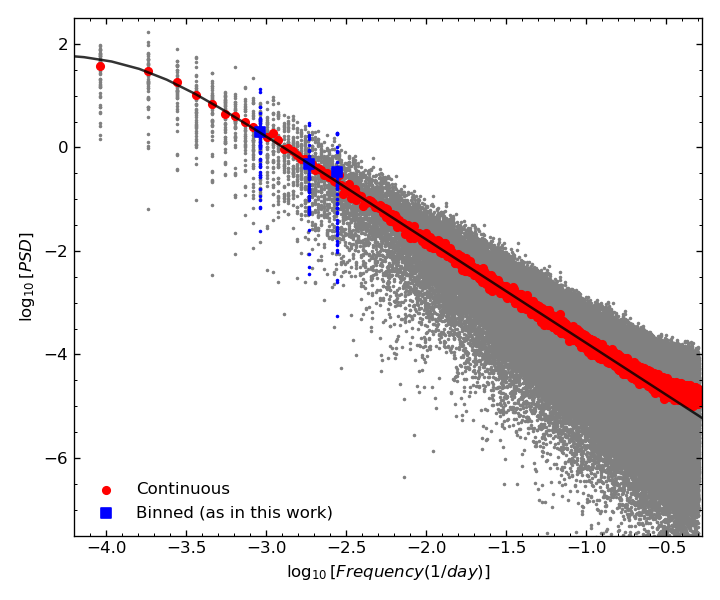}
    \caption{PSD of 50 simulated DRW light curves with a cadence of 1 day and a length of 20 years (grey points). Red points show the mean PSD values, at each frequency, on the original data. Blue squares are the ensemble PSD estimates from light curves binned as in this work. The black solid line shows the intrinsic PSD shape.}
    \label{fig:psd_sim}
\end{figure}

As an example, we show the resulting PSDs of simulations with $\tau = 10^3$ days in Fig. \ref{fig:psd_sim}. Small grey dots represent the periodograms of the 50 original light curves, while the red points indicate the mean periodogram at each frequency. It is clearly evident that the mean of the periodogram estimates is equal to the intrinsic power-law model at all frequencies (solid black line, Eq. \ref{eq:psd_drw}). This result is expected, as the periodogram is an unbiased estimator of the power spectral density function. Blue dots (and squares) on the same figure indicate the 50 periodograms (and their mean)  which are calculated using the binned light curves (which consist of only six points). Again, they follow the intrinsic model for the frequency range they are sampling, showing that our approach is robust.

The slight flattening in the continuous PSD at very high-frequencies (i.e. at the right-side of the figure) is likely due to aliasing, which is due to the fact that higher (non-sampled) frequencies fold back to lower frequencies. This might also be affecting the highest frequency point on the binned PSD, but for power-law-like power spectra with a slope steeper than $-1$ at high frequencies the effect is typically small. As power decreases logarithmically with frequency, the amount of power folded back to lower frequencies is small with respect to the observed value. The steeper the power-law slope, the lesser the effect, with the exact contribution depending on the intrinsic PSD shape and the sampled frequency range. Aliasing effects should also be suppressed as each one of the six points in the light curves is the result of binning the original light curves over discrete time intervals of a couple of months. This procedure significantly reduce the amount of power which is aliased back to the sample frequencies, as shown by \cite{vanderklis1988}.

Our simulations show that the worst case scenario should appear if we sample an intrinsic PSD with a slope flatter than $-1$ at high frequencies and a constant power at low frequencies (e.g. if the sampled frequencies are close the DRW break frequency). This could lead to underestimation in the recovered PSD slope up to $\sim 0.5$. However, this systematic can be easily reduced when more frequencies are available (i.e. with better sampled light curves), or corrected once a particular PSD model is assumed in order to explain the observed PSD slopes.

\end{appendix}

\end{document}